\documentstyle[epsf,psfig]{mn}
\begin{document}
\title{An excursion set model of hierarchical clustering: 
Ellipsoidal collapse and the moving barrier}
\author[R. K. Sheth \& G. Tormen]{Ravi K. Sheth$^1$ \& Giuseppe Tormen$^2$\\
$^1$ NASA/Fermilab Astrophysics Center, MS209, Batavia IL 60510-0500, USA\\
$^2$ Dipartimento di Astronomia, 35122 Padova, Italy \\
\smallskip
Email: sheth@fnal.gov, tormen@pd.astro.it
}
\date{Submitted to MNRAS 05 May 2001}

\maketitle

\begin{abstract}
The excursion set approach allows one to estimate the abundance and 
spatial distribution of virialized dark matter haloes efficiently and 
accurately.  The predictions of this approach depend on how the nonlinear 
processes of collapse and virialization are modelled.  We present simple 
analytic approximations which allow us to compare the excursion set 
predictions associated with spherical and ellipsoidal collapse.  
In particular, we present formulae for the universal unconditional mass 
function of bound objects and the conditional mass function which describes 
the mass function of the progenitors of haloes in a given mass range 
today.  We show that the ellipsoidal collapse based moving barrier model 
provides a better description of what we measure in the numerical 
simulations than the spherical collapse based constant barrier model, 
although the agreement between model and simulations is better at large 
lookback times.  Our results for the conditional mass function can be 
used to compute accurate approximations to the local-density mass function 
which quantifies the tendency for massive haloes to populate denser regions 
than less massive haloes.  This happens because low-density regions can be 
thought of as being collapsed haloes viewed at large lookback times, 
whereas high-density regions are collapsed haloes viewed at small lookback 
times.  Although we have only applied our analytic formulae to two simple 
barrier shapes, we show that they are, in fact, accurate for a wide variety 
of moving barriers.  We suggest how they can be used to study the case in 
which the initial dark matter distribution is not completely cold.  
\end{abstract}

\section{Introduction}
Bond et al. (1991) described an approach which allowed them to use 
the statistical properties of the initial density fluctuation field to 
derive an estimate of the number density of collapsed dark matter haloes 
at later times: the so-called universal, unconditional mass function 
(Press \& Schechter 1974).  
Lacey \& Cole (1993) showed how this model could be extended to estimate 
the rate at which small objects merge with each other to produce larger 
objects.  This allowed them to estimate the conditional mass function of 
subhaloes within parent haloes.  Lacey \& Cole also provided formulae 
for the distribution of halo formation times; Nusser \& Sheth (1999) 
provided formulae for the distribution of the halo mass at formation.  
Sheth (1996) and Sheth \& Pitman (1997) showed how various higher order 
statistical properties of the forest of merger history trees associated 
with the formation of these objects could also be estimated within this 
approach.  Mo \& White (1996) and Sheth \& Lemson (1999a) showed how 
information about the forest of merger history trees could be used to 
quantify the extent to which dark haloes are biased tracers of the 
matter distribution.  Mo, Jing \& White (1997) provided predictions 
for the higher order moments of the spatial distribution of the haloes, 
and Sheth (1998) showed how to use the approach to estimate the 
probability that a randomly placed cell contains a certain amount of 
mass.  Clearly, the approach has been very useful.  

The approach combines the simple physics of the spherical collapse model 
with the assumption that the initial fluctuations were Gaussian and small.  
The problem of estimating any one of the quantities listed above is 
reduced to solving a problem associated with the crossing of an 
appropriately chosen barrier by particles undergoing Brownian motion; 
the Brownian nature of the motion comes from the assumption of Gaussian 
initial conditions, and the barrier shape is specified by the spherical 
collapse model (e.g. Sheth 1998).  Hence, this is often called the 
excursion set approach.  

Tormen (1998) reported that the spherical collapse based excursion set 
predictions did not describe the conditional mass function of subclumps 
in his simulations well.  Motivated by this, Sheth, Mo \& Tormen (2001) 
discussed a simple way of modifying the excursion set approach to 
incorporate the effects of ellipsoidal, rather than spherical collapse.  
On average, initially denser regions collapse before less dense ones.  
This means that, at any given epoch, there is a critical density which 
must be exceeded if collapse is to occur.  In the spherical collapse 
model, this critical density does not depend on the mass of the collapsed 
object.  However, in their parametrization of ellipsoidal collapse, 
Sheth, Mo \& Tormen showed that, of the set of objects which collapse 
at the same time, the less massive ones must initially have been denser 
than the more massive ones, since the less massive ones would have had 
to hold themselves together against stronger tidal forces.  They argued 
that this could be incorporated into the excursion set approach, simply 
and effectively, if not rigorously, by changing the barrier shape.  
In essence, whereas the barrier associated with spherical collapse is 
one whose height does not depend on distance from the origin of the 
walk, the one associated with ellipsoidal dynamics increases with distance.  
They showed that the excursion set approach with a moving barrier was 
able to provide a good fit to the universal halo mass function.  

This paper is devoted to a more detailed discussion of moving barrier 
models.  In general, moving barrier models have a richer structure than 
the constant barrier model.  
For example, the approach with spherical dynamics predicts that, at any 
given time, all the mass in the universe is bound up in collapsed objects, 
whereas a small fraction of the mass remains unbound in the case of 
ellipsoidal dynamics.  In addition, whereas clustering is strictly 
hierarchical in the case of spherical dynamics, incorporating ellipsoidal 
collapse into the excursion set approach results in a model in which 
fragmentation as well as mergers may occur---the approach predicts that 
some small haloes fragment before they are subsumed into larger ones.  
Appendix~A describes some of these features, which may (or may not!) 
provide better approximations to the physics of graviational instability 
than does the constant barrier model, in more detail.  The Appendix 
also describes how the results of this paper can be used to model the 
halo mass function in warm dark matter scenarios such as that revisited 
by Bode, Ostriker \& Turok (2001).  

Section~\ref{barriers} shows how moving barrier models can be used to 
make simple analytic estimates of a number of statistical quantities
which are routinely measured in numerical cosmological simulations.  
The primary results of Section~\ref{barriers} are equations~(\ref{taylors}) 
and~(\ref{fsS}), which are accurate for a large class of moving barrier 
shapes.  To illustrate how these formulae work, we show the result of 
inserting the ellipsoidal collapse moving barrier of 
Sheth, Mo \& Tormen (2001) into these formulae.  Section~\ref{umf} 
presents the number density of bound objects as a function of mass 
(the unconditional mass function), and Section~\ref{generate} describes 
a simple efficient algorithm for generating it.  
Section~\ref{cmf} presents the average 
number of progenitor subhaloes as a function of subhalo mass, for a wide 
range of specified parent halo masses (the conditional mass function), 
and Section~\ref{dmf} shows how the halo mass function depends on the 
surrounding density field (the local-density mass function).  
Comparison with simulations shows that the approach, with the 
ellipsoidal collapse based moving barrier shape, is quite accurate.  
Section~\ref{rescale} shows that, at small lookback times, neither the 
constant nor the moving barrier predictions describe the conditional 
mass functions in the simulations particularly well, although the 
agreement at large lookback times is quite good if the spherical 
collapse constant barrier is used, and even better if the ellipsoidal 
collapse moving barrier is used.  

Section~\ref{sixd} shows the result of considering more complicated moving 
barrier excursion set models.  In particular, it shows the result of 
considering the full six-dimensional random walk associated with Gaussian 
random fields, rather than the one-dimensional simplification proposed by 
Sheth, Mo \& Tormen (2001).  It shows that their simplification is 
actually quite accurate.  Details associated with the calculations in 
this section are presented in Appendix~\ref{grfs}.  

Section~\ref{discuss} discusses some simple implications of our findings.  
Although, in this paper, we concentrate on the moving barrier derived by 
Sheth, Mo \& Tormen (2001), we think it worth stressing that our analytic 
formulae are more general:  they are accurate for a wide variety of moving 
barrier shapes.  

\section{The moving barrier model}\label{barriers}
As discussed in the introduction, we will mainly be interested in 
the first crossing distributions of uncorrelated Brownian motion 
random walks.  Following Bond et al. (1991) and Lacey \& Cole (1993), 
these first crossing distributions can be used to provide useful 
approximations to what have come to be called the conditional and 
unconditional mass functions of the dark halo distribution.  
The results of this section should be thought of as generalizations 
of the results in Lacey \& Cole (1993).  Whereas they restricted their 
attention to a barrier of fixed height, this section presents analytic 
formulae which approximate the barrier crossing distribution for a wide 
class of moving barriers.  

Before we begin, we think a word on notation is helpful.  
We have chosen to present our formulae for the first crossing 
distributions using the same notation as Lacey and Cole (1993).  
This means that we use the symbol $S$ to represent the variance in the 
density fluctuation field when smoothed on some scale, which is usually 
denoted $\sigma^2$.  However, some of our formulae can be written 
in terms of the scaled variable $(\sigma/\sigma_*)^2\equiv S/S_*$, 
for some suitably defined $S_*$.  When this is possible, we also write 
our formulae in terms of $\nu\equiv S_*/S$.

\subsection{The unconditional mass function}\label{umf}
To illustrate how our formulae work, rather than use the spherical collapse 
barrier of constant height, we will use the moving barrier shape derived by 
Sheth, Mo \& Tormen (2001)---the one associated with ellipsoidal collapse:  
\begin{equation}
B(\sigma^2,z) = \sqrt{a}\,\delta_{\rm sc}(z) 
\bigl[1 + \beta\,(a\,\nu)^{-\alpha} \Bigr],
\label{bec}
\end{equation}
where $\nu\equiv [\delta_{\rm sc}(z)/\sigma(m)]^2$.  
In the expression above, $\delta_{\rm sc}(z)$ is the critical overdensity 
required for spherical collapse at $z$, extrapolated using linear theory 
to the present time (it is approximately 1.686$(1+z)$ if $\Omega =1$), 
and $\sigma(m)$ is the rms of the initial density fluctuation field 
when it is smoothed on a scale which contains mass $m$, extrapolated 
using linear theory to the present time.  
The parameters $\beta\approx 0.485$ and $\alpha\approx 0.615$ come from 
ellipsoidal dynamics (the spherical collapse model has $\alpha=0$ and 
$\beta=0$), and the value $a\approx 0.7$ comes from normalizing 
the model to simulations (as we discuss later, it may be more accurate 
to set $a\approx 0.75$ or so).  The spherical collapse model sets 
$B(\sigma^2,z) = \delta_{\rm sc}(z)$; because $B$ is then independent 
of $\sigma$, spherical collapse is said to be associated with a barrier 
of constant height.  

As Sheth, Mo \& Tormen noted, $B(\sigma^2,z)$ scales similarly to the 
natural scaling associated with random walks:  multiplying the barrier 
height by $c$ and rescaling $\sigma$ by the same factor results 
in a barrier of the same shape.  They exploited this property as follows:  
first, they studied what happens when $z=0$.  Whereas an analytic 
expression for the first crossing distribution of a barrier of constant 
height has been known for some one hundred years or so, there is as yet 
no analogous solution for the ellipsoidal collapse moving barrier.  
(One of the main results of this section is to provide a good analytic 
approximation to this solution.)  
So, Sheth, Mo \& Tormen simulated a large number of random walks and 
computed the distribution of first crossings, $f(S)$, of the 
barrier $B(\sigma^2=S,z=0)$ numerically.  The scaling of the barrier 
shape meant that their numerical solution could be scaled to represent
the first crossing distribution at any other time also.  Therefore, 
when providing an analytic fit to their simulated first crossing 
distribution, they expressed it in scaled variables:  
\begin{equation}
\nu\,f(\nu) = A\,\Bigl[1 + (a\,\nu)^{-p}\Bigr]\ 
\left({a\,\nu\over 2}\right)^{1/2}\,{{\rm e}^{-a\nu/2}\over\sqrt{\pi}} ,
\label{giffit}
\end{equation}
where $f$ is the distribution of first crossings, $\nu$ and $a$ are the 
variables defined above, $p=0.3$ and $A$ is determined by requiring that 
the integral of $f(\nu)$ over all $\nu$ give unity.  The distribution 
associated with the spherical collapse constant barrier is got by 
setting $a=1$, $p=0$ and $A=1/2$.  

In the excursion set approach, the average comoving number density of halos 
of mass $m$, often called the universal or unconditional halo mass function 
$n(m,z)$, is related to the first crossing distribution (this is why
the shape of the barrier influences the shape of the mass function) by 
\begin{equation}
\nu\,f(\nu)\equiv m^2\,{n(m,z)\over \rho}\,
{{\rm d}\,\ln m\over {\rm d}\,\ln\nu},
\label{fnunm}
\end{equation}
where $\rho$ denotes the average comoving density of the background.  
Because the first crossing distribution evolves self-similarly, 
so does the halo mass function.  
Fig.~2 in Sheth \& Tormen (2001) shows that, in the GIF simulations
of clustering in SCDM, OCDM and $\Lambda$CDM cosmologies, $n(m,z)$
is well fit by this expression (at least over the range $z=0$ to $z=4$).  
That is to say, the excursion set prediction that the universal 
unconditional mass function evolves self-similarly, is actually a 
very good approximation.

Although this fitting function (equation~\ref{giffit}) for the first 
crossing distribution is extremely useful, when we consider conditional 
mass functions in the next section, it will turn out that we need to 
compute a new fitting function for each parent mass range of interest.  
This is, in principle, a computational bottleneck.  For this reason, 
we think it more useful to provide a formula for the first crossing 
distribution which is more amenable for use in what is to follow.  

We have analytic formulae for the first crossing distribution only in 
the case of constant (e.g. Bond et al. 1991) and linear (Sheth 1998 and 
Appendix~A of this paper) barriers.  These solutions suggest the 
following approximation which we have found to work rather well.  For 
a wide range of moving barrier shapes $B(S)$, the first crossing 
distribution is well approximated by 
\begin{equation}
f(S)\,{\rm d}S = |T(S)|\,\exp\left(-{B(S)^2\over 2S}\right)\,
{{\rm d}S/S\over\sqrt{2\pi S}},
\label{taylors}
\end{equation}
where $T(S)$ denotes the sum of the first few terms in the Taylor 
series expansion of $B(S)$:  
\begin{displaymath}
T(S) = \sum_{n=0}^5 {(-S)^n\over n!} {\partial^n B(S)\over\partial S^n}.
\end{displaymath}
Notice that this expression gives the correct answer in the case of 
constant and linear barriers (in which only the first, or the first two 
terms of the series are non-zero).  For the ellipsoidal barrier shape, 
we find we get reasonable accuracy to the numerical result if we truncate 
this expansion at $n=5$.  The accuracy of this formula increases as the 
distance between the start of the walk and the barrier height at that 
initial position, increases (this distance is $\sqrt{a}$ for the 
ellipsoidal collapse barrier of equation~\ref{bec}).  

\begin{figure}
\centering
\epsfxsize=\hsize\epsffile{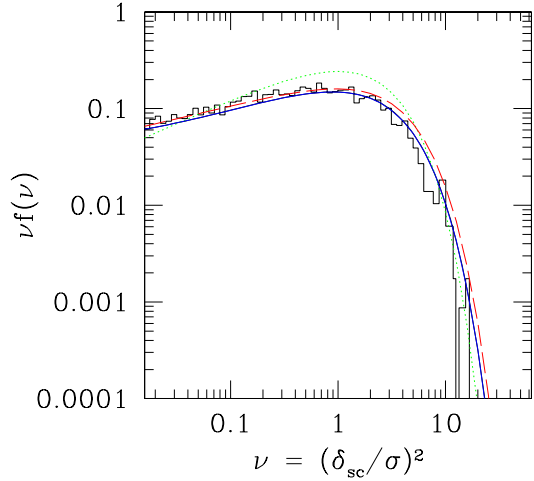}
\caption{First crossing distributions and the universal 
unconditional halo mass function.  Histogram shows the distribution 
obtained by simulating random walks which are absorbed on the 
ellipsoidal collapse moving barrier; solid curve shows our analytic 
approximation to this distribution (equation~\ref{taylors}).  
Dashed curve shows the distribution which fits the halo mass function 
in cosmological simulations (equation~\ref{giffit}), and dotted curve 
shows the distribution associated with spherical collapse.  }
\label{umfcomp}
\end{figure}

To illustrate, Fig.~\ref{umfcomp} shows the result of setting the 
barrier $B(S)$ to equal that given by equation~(\ref{bec}) at $z=0$, 
then generating a large ensemble of random walks, and so constructing 
the first crossing distribution.  The histogram shows this Monte-Carlo 
distribution, and the solid curve shows the approximation presented in 
equation~(\ref{taylors}):
\begin{eqnarray}
 \nu\,f(\nu) &=& \sqrt{{a\nu\over2\pi}}
               {\rm e}^{-a\nu [1 + \beta (a\nu)^{-\alpha}]^2/2}
               \Big(1 + \nonumber \\
              && \ \ {\beta\over (a\nu)^\alpha}
\Bigl[1 - \alpha + {\alpha(\alpha-1)\over 2}+\cdots\Bigr]\Big)\nonumber \\
             &\approx& \Big(1 + {0.094\over (a\nu)^{0.6}}\Big) 
               \sqrt{{a\nu\over2\pi}}
               {\rm e}^{-a\nu [1 + 0.5 (a\nu)^{-0.6}]^2/2}.
\end{eqnarray}
These should be compared with the distribution of first crossings of 
a barrier of constant height set equal to $\delta_{\rm sc}(z=0)$ 
(dotted curve), and equation~(\ref{giffit}) which fits the GIF 
simulations (dashed curve).  
With the exception of the dotted curve (the one associated with the 
constant barrier spherical collapse model) the other three curves are 
in reasonably good agreement.  We will exploit this fact in the next 
subsection, when we derive a simple expression for the conditional 
mass function associated with ellipsoidal collapse.

Before we do so, however, we think it useful to point out that 
there are some generic properties of moving barrier models which arise 
from the scaling of the barrier shape.  The fact that the barrier shape 
scales in the way it does means that moving barrier models in which the 
barrier height increases with $S$ are somewhat more complicated than 
the constant barrier model, so they may be used to model a wider variety 
of physical processes.  For example, our final expression for 
the moving barrier crossing distribution is not normalized to unity 
(notice that the solid curve is always below the dashed one).  This 
is because, if the barrier height increases more steeply than linearly 
with $\sigma$ then not all walks cross the barrier.  
Appendix~A uses a simple analytic moving barrier model to illustrate 
some of these features.  

\subsection{Generating the unconditional mass function}\label{generate}
Before we move on to study the conditional mass function, we thought 
it useful to describe a fast and simple algorithm for generating random 
numbers which can be used to provide the correct distribution of halo 
masses.  To generate numbers drawn from the spherical collapse mass 
function (equation~\ref{giffit} with $A=1/2$ and $p=0$) is easy
because, in this case, $f(\nu)$ can be got by generating a Chi-squared 
random variate.  This means that, in the case of spherical collapse, 
the unconditional mass function can be generated quickly and easily 
because generating Gaussian random variates is easy.  In particular, 
one simply generates a Gaussian random variate $x$, and then sets 
$a\nu = x^2$.  

The ellipsoidal collapse mass function (equation~\ref{giffit} with 
$A=0.32218$ and $p=0.3$) cannot be transformed to a Gaussian, so 
constructing an algorithm for generating it is not so straightforward.  
We have found that first generating a Gaussian variate $x$, and then 
setting $a\nu = |x|^{3.6}/(1 + |x|^{1.6})$ is accurate to within a 
percent or so over the range $0.01 \le \nu\le 100$.  
The speed of this algorithm compensates for the fact that it does 
not exactly produce variates drawn from the ellipsoidal collapse 
mass function.  

\subsection{The conditional mass function}\label{cmf}
As stated above, we do not know of an analytic expression for the 
first crossing distribution associated with barriers which have the 
form given in equation~(\ref{bec}).  However, we do have two 
reasonably accurate fitting formulae---equations~(\ref{giffit}) 
and~(\ref{taylors})---to this distribution.  
One might have thought that we could use them to make an estimate 
of the conditional mass function as follows.   

Bond et al. (1991) and Lacey \& Cole (1993) argued that conditional mass 
functions could be estimated by considering the successive crossings of 
boundaries associated with different redshifts.  The first crossing of 
two constant barriers of different heights has an analytic solution, so
they were able to provide analytic estimates for the conditional mass 
function associated with the spherical collapse model.  
Such a formula is very useful, because, once the conditional 
mass function is known, the forest of merger history trees can be 
constructed using the algorithm described by Sheth \& Lemson (1999b), 
from which the nonlinear stochastic biasing associated with this mass 
function can be derived using the logic of Mo \& White (1996) and 
Sheth \& Lemson (1999a).  

\begin{figure}
\centering
\mbox{\psfig{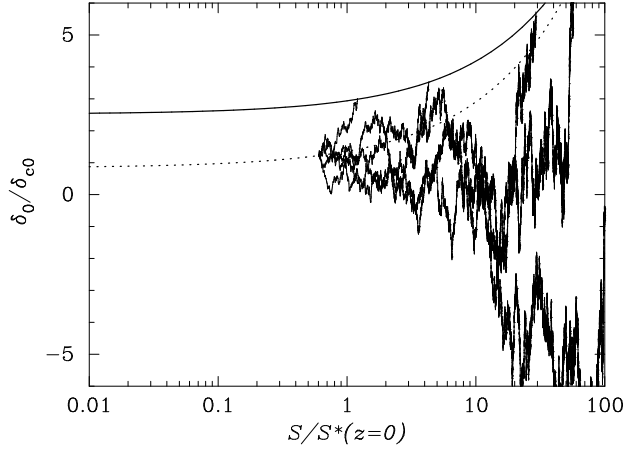}}
\caption{Examples of random walks used to construct the conditional 
mass functions associated with the ellipsoidal collapse moving 
barriers at $z=0$ (dotted curve) and $z=2$ (solid curve).  }
\label{gif2bar}
\end{figure}

In the constant barrier case, the conditional mass function is 
computed by considering walks which start from 
$[\sigma^2=S,\delta_{\rm sc}(z_0)]$ rather than from the origin, 
and then intersect the constant barrier $\delta_{\rm sc}(z_1)$ at, 
say, $s$.  This is easily computed, because, despite the shift in the 
origin, the second barrier is still one of constant height.  Since the 
first crossing distribution of a barrier of constant height is known, 
(recall it was just such a distribution which was associated with 
the universal, unconditional mass function), the conditional mass 
function can also be written analytically.  Essentially, $f(s|S)$ 
has the same form as the unconditional mass function $f(S)$, but 
with the change of variables:  
$\delta_{\rm sc}(z)\to \delta_{\rm sc}(z_1)-\delta_{\rm sc}(z_0)$ 
and $S\to s-S$.  

One might have wondered if the same change of variables in 
equation~(\ref{giffit}) provides a good description of the conditional 
mass function associated with the ellipsoidal collapse moving barrier.  
Unfortunately, because the barrier shape is not linear in $S$, changing 
the origin of the coordinate system does not yield a barrier of 
exactly the same functional form.  Specifically, the shape of 
\begin{eqnarray*}
B(s,z_1) - B(S,z_0) &=& 
\sqrt{a}\,\delta_1\bigl[1 + \beta\,s^\alpha/(a\,\delta_1^2)^\alpha \Bigr] 
\nonumber\\
&& 
- \sqrt{a}\,\delta_0\bigl[1 + \beta\,S^\alpha/(a\,\delta_0^2)^\alpha \Bigr] ,
\end{eqnarray*}
where $\delta_1 \equiv \delta(z_1)$ etc., 
can be written as a constant plus a term which scales as $(s-S)^\alpha$ 
only when $\alpha$ equals zero or one.  This means that, formally, the 
solution to the two barrier problem associated with ellipsoidal dynamics 
is not given by a simple rescaling of the unconditional ellipsoidal 
collapse mass function.  Therefore, we cannot simply rescale the fitting 
function of equation~(\ref{giffit}) to get a reliable estimate of the 
conditional mass function:  the two barrier problem associated with 
moving barriers must, in general, be solved numerically.  

This is discouraging because it means that, in principle, we must find 
a different fitting function for each choice of condition, because 
each condition corresponds to a different origin, say, $(B_0,S_0)$, 
and so to a slightly different barrier shape.  
Of course, the result can be generated relatively quickly in at least 
two ways.  The first is to solve the integral equation associated with 
this barrier numerically (Monaco 1997b; Sheth 1998).  
The second is to simply simulate the random walk trajectories and so 
construct the first crossing distribution directly.  

Fig.~\ref{gif2bar} shows an example of what is involved in solving 
the two barrier problem numerically in this way.  The smooth dotted 
and solid curves show the ellipsoidal collapse barrier $B(S,z)$ of 
equation~(\ref{bec}) scaled to $z=0$ and $z=2$, respectively.  
Jagged curves show a few representative random walk trajectories:  
they start at the barrier position $B(S,z=0)$, where $S(M)$ is given 
by the GIF SCDM power spectrum, and $M/M_*=2$.  These random walks are 
followed until they first cross the barrier $B(s,z=2)$.  The value of 
$S$ at which this happens is stored and used to make plots like those 
shown below.  
The random walks were generated by the same Monte--Carlo code that 
was used to generate Fig.~\ref{lin2bar}, except that there the barrier 
shape was given by equation~(\ref{linbar}).  Fig.~\ref{lin2bar} shows 
that this Monte--Carlo code works correctly.  
One possible approach to the excursion set conditional mass function, 
then, is to simply generate it numerically, as the need arises.  

\begin{figure*}
\centering
\epsfxsize=\hsize\epsffile{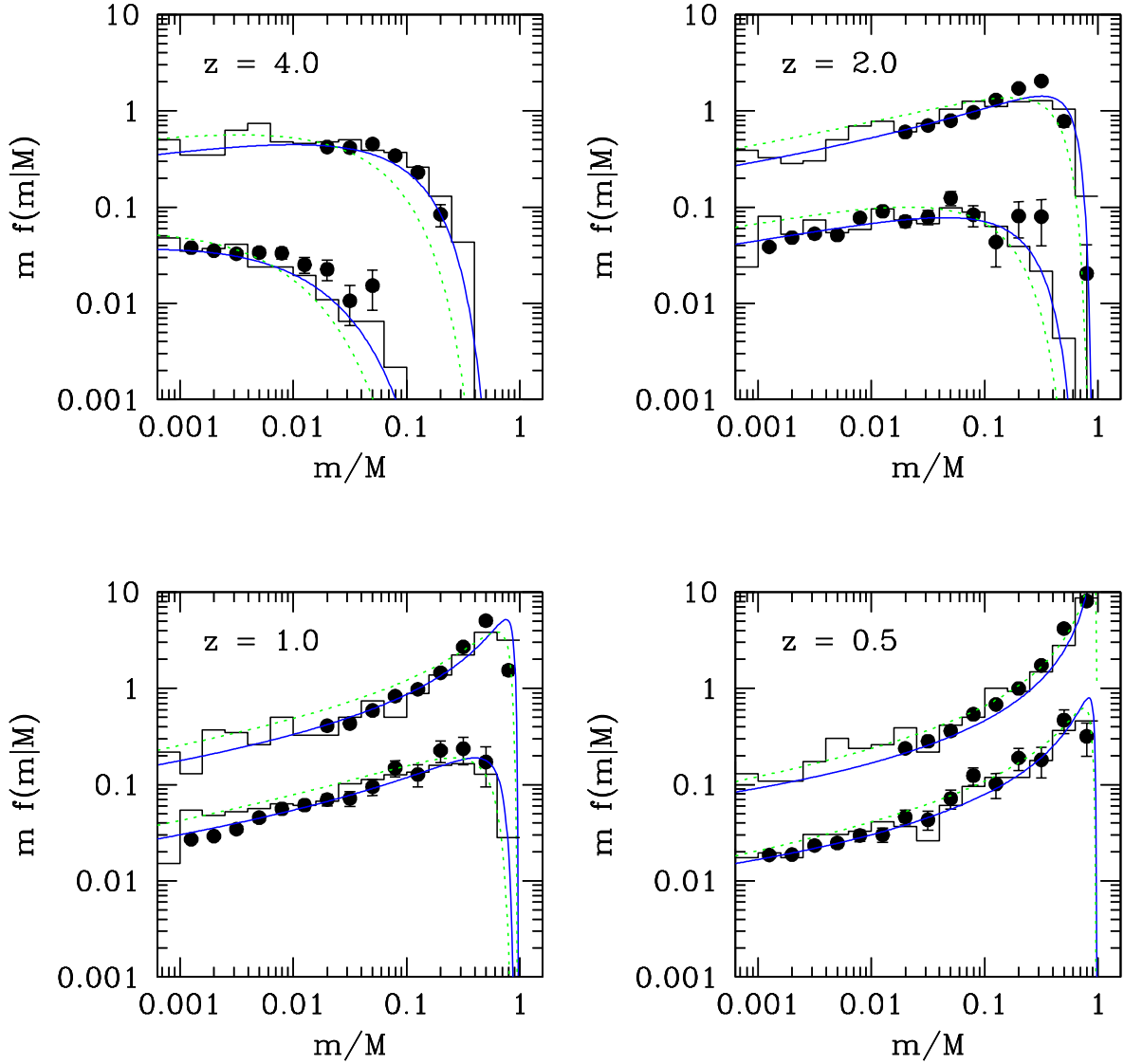}
\caption{Excursion set conditional mass functions at $z$, for parent 
haloes identified at the present time.  The two sets of curves in each 
panel are for parent haloes in the mass range $1\le M/M_*\le  2$ 
(upper curves) and $16\le M/M_*\le  32$ (lower curves); the upper 
curves have been shifted upwards by a factor of ten for clarity.  
Symbols with error bars show the distributions measured in the 
$\Lambda$CDM simulations, histograms show the result of generating 
the first crossing distribution by simulating an ensemble of $10^4$ 
random walks, smooth solid curves show the analytic approximation 
discussed in the text, and dotted curves show the distribution associated 
with barriers of constant height.  } 
\label{clcdm}
\end{figure*}

Before providing a detailed comparison of the conditional mass 
functions generated using this Monte--Carlo model and those in 
numerical simulations, it is useful to study a simple limiting case.  
Fig.~\ref{gif2bar} shows that for $S/S_*<0.5$, the height of the 
barrier is approximately constant.  At small $S$, the only 
difference between the barrier at two redshifts, and the spherical 
collapse constant barriers, arises from the factor of $a=0.707$.  
This has the following consequence.  At small lookback times 
(small redshift differences), most random walk trajectories will 
intersect the barrier before they have travelled very far along 
the $S$ axis.  For these trajectories, the barrier is, effectively 
one of constant height.  This means that the conditional mass function 
for massive haloes at small lookback times will have the same shape as 
that predicted by the constant barrier, with one small difference.  
The factor of $a=0.707$ has the effect of slightly reducing (by a factor 
of $\sqrt{a}$) the effective redshift difference relative to the original 
constant barrier model.  As a result, the GIF barrier suggests that 
massive haloes at small lookback times will be slightly more massive 
than the original constant barrier model predicts.  Since $\sqrt{0.707}$ 
is close to unity, this effect will be small.  In practice, we only expect 
the barrier predictions to differ significantly from those of the 
constant barrier for small haloes, or at large lookback times 
(high redshift).  This is encouraging, because these are precisely 
the regimes in which simulations suggest that the constant barrier 
model is inaccurate (Tormen 1998).  

\begin{figure*}
\centering
\epsfxsize=\hsize\epsffile{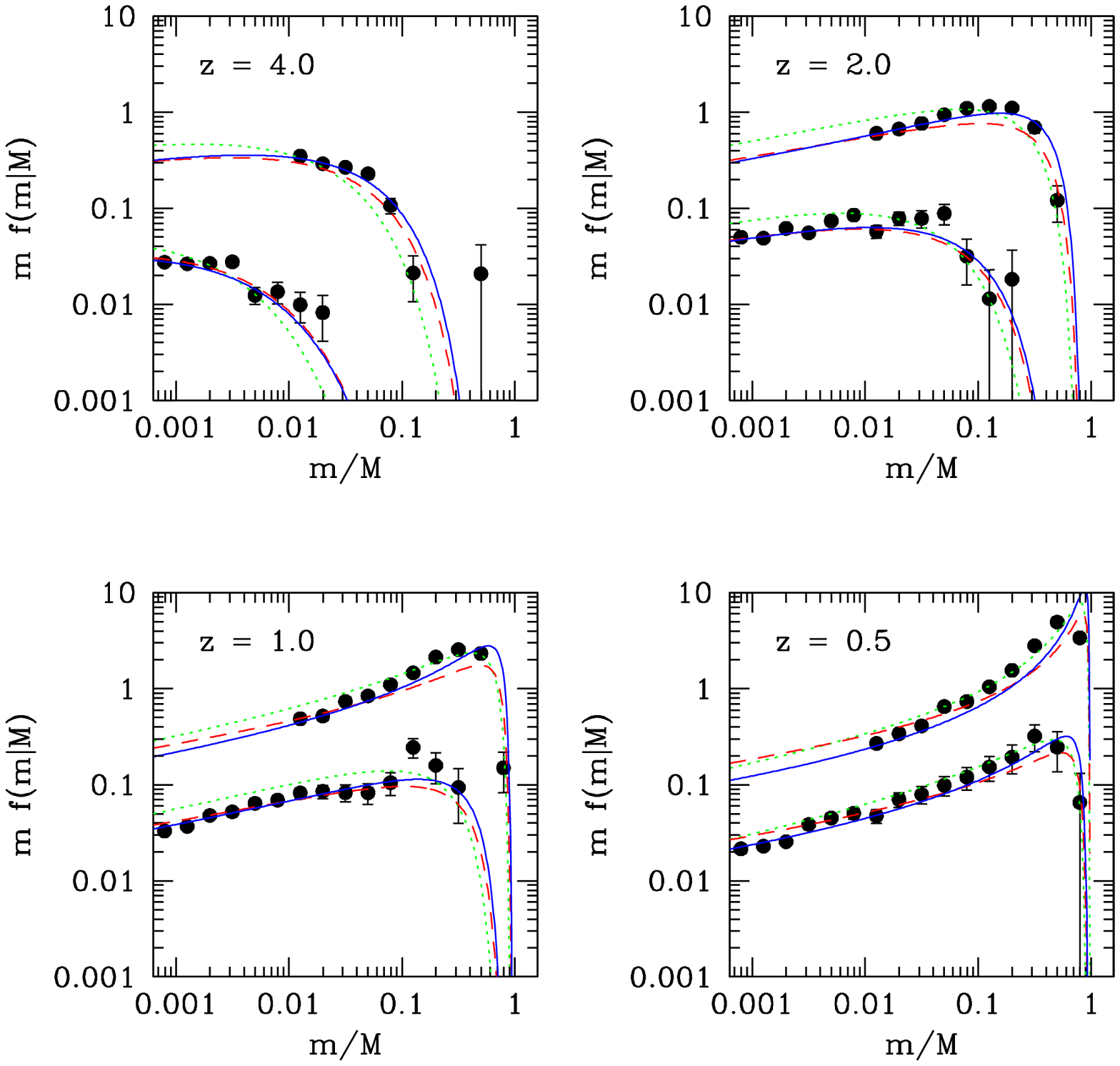}
\caption{Conditional mass functions in the SCDM simulations (symbols 
with error bars).  The two sets of curves in each panel show parent 
haloes with masses in the range $1\le M/M_*\le  2$ (upper curves) and 
$16\le M/M_*\le  32$ (lower curves); the upper curves have been shifted 
upwards by a factor of ten for clarity.
Dotted curves show the spherical collapse prediction (Lacey \& Cole 1993), 
dashed curves show the distribution one gets by rescaling the 
unconditional mass function (equation~\ref{giffit}), 
and solid curves show our analytic approximation to the random walk 
with moving-barrier simulations (equation~\ref{fsS}).}
\label{cscdm}
\end{figure*}

In addition to Monte-Carloing the conditional mass functions, we can 
use the results of the previous subsection to derive a simple analytic 
expression for their shape.  We can do this because our barrier crossing 
formula is reasonably accurate for a rather wide range of barrier shapes.  
A glance at Fig.~\ref{gif2bar} shows that the barrier shapes associated 
with the conditional mass functions are not likely to be too different 
from those associated with the unconditional function, so we should be 
able to use equation~(\ref{taylors}) to approximate most of the 
conditional mass functions we will be interested in.  
In practice, this can be done by simply making the appropriate 
replacements $B\to B(s)-B(S)$ and $S\to s-S$ in equation~(\ref{taylors}).  
At the risk of being repetitive, our approximation for the conditional 
mass function is:
\begin{equation}
{\cal N}(m,\delta_1|M,\delta_0)\,{\rm d}m \equiv 
{M\over m}\,f(m,\delta_1|M,\delta_0)\,{\rm d}m
\label{NmM}
\end{equation}
where $f(m|M)\,{\rm d}m = f(s|S)\,{\rm d}s$ with 
\begin{equation}
f(s|S)\,{\rm d}s = {|T(s|S)|\over \sqrt{2\pi (s-S)}}\,
\exp\left(-{[B(s)-B(S)]^2\over 2(s-S)}\right)\,{{\rm d}s\over s-S},
\label{fsS}
\end{equation}
and 
\begin{displaymath}
T(s|S) = \sum_{n=0}^5 {(S-s)^n\over n!} 
{\partial^n [B(s)-B(S)]\over\partial s^n}.
\end{displaymath}

Fig.~\ref{clcdm} compares this approximation with the actual 
numerical Monte-Carlo distribution, and compares both with the 
actual distribution measured in the cosmological simulations.  
(We will show a comparison with what one gets by rescaling 
equation~\ref{giffit} shortly.)  The histograms (without error bars) show 
the conditional mass functions generated using our Monte--Carlo code, 
for parent haloes in the mass range $1\le M/M_*\le  2$ (upper curves) 
and $16\le M/M_*\le  32$ (lower curves); the upper curves have been shifted 
upwards by a factor of ten for clarity.  The smooth curves show the 
associated analytic approximation; they describe the results of our 
numerical walks reasonably well.  Therefore, in the comparisons to follow, 
we will sometimes only show the analytic curves.  Recall that the analytic 
formula should be most accurate for the high-redshift progenitors of 
massive parents, and least accurate when the parents are not very 
massive.  The figure shows some evidence that this is true.  
Both these curves should be compared with the symbols which have 
error bars---they show the conditional mass functions measured in the 
$\Lambda$CDM simulations.  The figure is quite encouraging: our excursion 
set predictions are in reasonably good agreement with the cosmological 
simulations.  In addition, the figure shows that the analytic 
approximation (equation~\ref{fsS}) is reasonably accurate.  

Fig.~\ref{cscdm} shows a similar comparison in the case of SCDM.  
In this case, we have chosen to not show the random walk histogram, 
since it is quite well fit by our formula.  Instead, the figure 
shows the conditional mass functions measured in the SCDM simulations 
(symbols with error bars) and the various analytic approximations 
discussed earlier (smooth curves).  
In order of accuracy, these are equation~(\ref{fsS}) (solid), 
the distribution associated with rescaling the unconditional mass 
function of equation~(\ref{giffit}) (dashed), and the conditional 
distribution associated with the constant barrier, spherical collapse 
model of Lacey \& Cole (1993) (dotted).  (The finite mass resolution 
of the simulations means that the distributions are artificially 
truncated at low masses.)

Notice that the spherical collapse based dotted curves are often 
quite different from the N-body simulation symbols.  This discrepancy 
is similar to that first noticed by Tormen (1998); haloes in the 
simulations seem to hold themselves together at higher redshift than 
the spherical collapse model predicts.  Notice also that, whereas the 
dashed curves we get by rescaling the unconditional halo mass function 
are certainly better fits to the cosmological simulations, the solid 
curves, in which the relation between the excursion set model and the 
conditional mass function are accounted for more carefully, are almost 
always even more accurate.  

\subsection{Dependence on local density}\label{dmf}

\begin{figure}
\centering
\epsfxsize=\hsize\epsffile{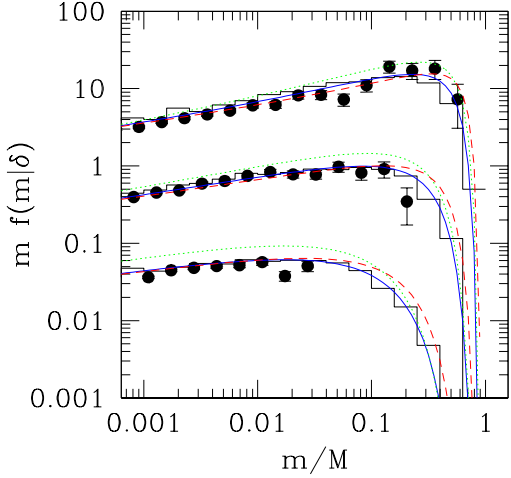}
\caption{Mass functions as a function of local density in the SCDM 
simulations (symbols with error bars), plotted as a function of 
relative mass, so as to resemble the conditional mass functions 
presented earlier.  Dotted curves show the spherical collapse prediction, 
dashed curves show the distribution one gets by rescaling the 
unconditional mass function (equation~\ref{giffit}), 
and solid curves show our analytic approximation to the random walk 
with moving-barrier simulations (equation~\ref{fsS}).  
The curves have been offset upwards by a factor of ten and a hundred, 
in the case of the middle and topmost curves, respectively.  
The upper most curves show the densest cells.}
\label{fdscdm}
\end{figure}

Following Mo \& White (1996), knowledge of the conditional mass 
function allows one to estimate how the distribution of dark matter
haloes today depends on the average density in which the haloes are.  
In essence, they argued that a dense region should be thought of as 
an object which will collapse and form a virialized halo in the near 
future.  This means that the haloes in it today can be thought of as 
`progenitor subhaloes' viewed at `low redshift'.  
In contrast, it will be a much longer time before an underdense region 
collapses (if it collapses at all), so the haloes within it today are 
like the progenitor haloes viewed at high redshift.  
In hierarchical models, massive haloes form later, and less massive 
haloes form earlier.  The discussion above means that one expects 
haloes in dense regions to be more massive, on average, than in 
underdense regions.  The precise dependence of halo mass on local 
density depends on the precise relation between the local density 
today and the effective `redshift'.  Mo \& White used the spherical 
collapse model to provide this relation.  They provided a simple 
fitting formula to this relation in an Einstein-deSitter universe.  
Lemson \& Kauffmann (1999) and Sheth \& Lemson (1999a) showed that this 
provided a reasonably good description of how, in their simulations, 
the density of haloes depended on local density.  
We have checked that the following simple modification to their 
formula is reasonably accurate for all cosmologies of interest:  
\begin{eqnarray}
\delta_0(\delta,z_0)\!\!\! &=&\!\!\! {\delta_{\rm sc}(z_0)\over 1.68647}
 \times \Biggl[1.68647 - {1.35\over (1+\delta)^{2/3}} \nonumber\\
&&\qquad\qquad\quad - {1.12431\over (1+\delta)^{1/2}} 
                  + {0.78785\over (1+\delta)^{0.58661}}\Biggr],
\label{d0mow}
\end{eqnarray}
where $M/\bar\rho V\equiv (1+\delta)$ is the nonlinear density of a 
region containing mass $M$ within the volume $V$ at $z_0$, and 
$\delta_{\rm sc}(z_0)$ denotes the critical density for spherical 
collapse at $z_0$.  The number density of $m$-haloes in regions of 
nonlinear density $\delta$ is got from the conditional mass function 
${\cal N}[m,\delta_{\rm sc}(z_1)|M,\delta_0(\delta,z_0)]$ of 
equation~(\ref{NmM}).  

\begin{figure}
\centering
\epsfxsize=\hsize\epsffile{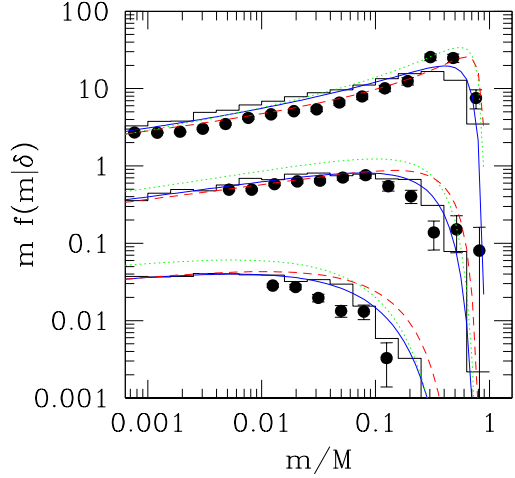}
\caption{Same as the previous figure, but for $\Lambda$CDM.}
\label{fdlcdm}
\end{figure}

The previous section showed that, in fact, the conditional mass 
functions are better fit by ellipsoidal collapse based curves.  
Therefore, one might reasonably expect that the same will be true 
for $n(m|\delta)$.  To emphasize how similar $n(m|\delta)$ is to 
the conditional mass function, we have chosen to do the following.
We have divided the simulation volume up into into cubes, each 
$10$Mpc/$h$ on a side.  We then divided the cubes into three classes:  
the densest, and least dense ten percent of the cells, and the ten 
percent around the median density.  
Figs.~\ref{fdscdm} and~\ref{fdlcdm} show $m^2\,n(m|\delta)$ for the 
haloes in the cells, plotted as a fraction of the mass of a cell.  
Symbols with error bars show the measurements in the cosmological 
simulations, histograms show the associated random walk distributions, 
solid curves show our analytic approximation to the random walks, 
dashed curves show the result one gets by rescaling the unconditional 
mass function, and dotted curves are for the spherical collapse, 
constant barrier model.  The curves for the three types of cells 
have been offset from each other for clarity; the lowest density 
cells are the lowest curves, and they have not been shifted, 
average density cells have been shifted upwards by a factor of ten, 
and the densest cells have been shifted upwards by a factor of a 
hundred. In all cases, the random walk model provides the best fit 
to the simulation data.  

Notice how similar these curves appear to the conditional mass functions 
presented earlier.  Because most of haloes in the densest cells are a 
significant fraction of the total mass in the cell, the mass function 
in dense cells looks very like the low redshift conditional mass 
functions.  In contrast, the mass function in underdense cells looks 
much more like the high redshift conditional mass functions.  This is 
precisely what the model predicts.  

\begin{figure}
\centering
\epsfxsize=\hsize\epsffile{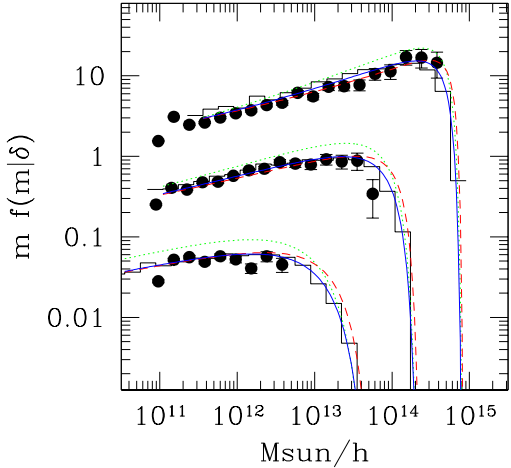}
\caption{Same as Fig.~\ref{fdscdm}, but now presented as a function 
of physical, rather than relative, mass.  Massive haloes occupy 
the densest cells (upper most curves).  }
\label{ndscdm}
\end{figure}

\begin{figure}
\centering
\epsfxsize=\hsize\epsffile{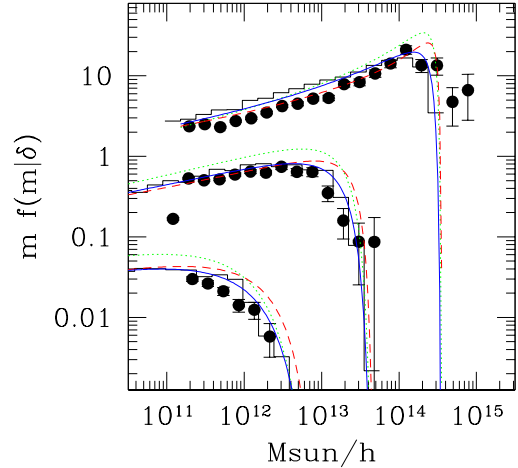}
\caption{Same as the previous figure, but for $\Lambda$CDM.}
\label{ndlcdm}
\end{figure}

Figs.~\ref{ndscdm} and~\ref{ndlcdm} show what this trend means for 
the actual number density of haloes in dense and less dense regions.  
The various symbols and curves are the same as in the previous two 
figures; the only difference is that now the x-axes have been multiplied 
by the total mass in the cell to show physical, rather than relative 
masses.  Clearly, less dense cells have essentially no massive haloes.  
In addition, the ratio of massive to less massive haloes is higher in 
denser cells.  In particular, note that the density of less massive 
haloes in dense regions is actually smaller than the density of less 
massive haloes in underdense regions.  One might have thought that 
dense regions simply have more haloes on average.  For example, 
one might have thought that $n(m|\delta) = (1+\delta)\,n(m)$.  
Our figures show that this is wrong, but that a good estimate of 
$n(m|\delta)$ can, nevertheless, be computed analytically.  We 
conclude this section with the observation that our moving barrier 
based formulae provide a more accurate fit to the simulations than 
does the spherical collapse based constant barrier model.  

\subsection{Rescaling the conditional mass function}\label{rescale}

In the excursion set model with a constant barrier height, the 
unconditional mass function, when expressed as a function of 
$\nu\equiv \delta^2_{\rm c}/\sigma^2(m)$, is expected to be a universal 
function which is independent of redshift, cosmology or initial 
power spectrum.  In addition, if the conditional mass function is 
expressed in units of 
 $(\delta_{\rm c1}-\delta_{\rm c0})^2/(s-S)$, then it is expected 
to have the same shape as the unconditional mass function.  

\begin{figure*}
\centering
\epsfxsize=\hsize\epsffile{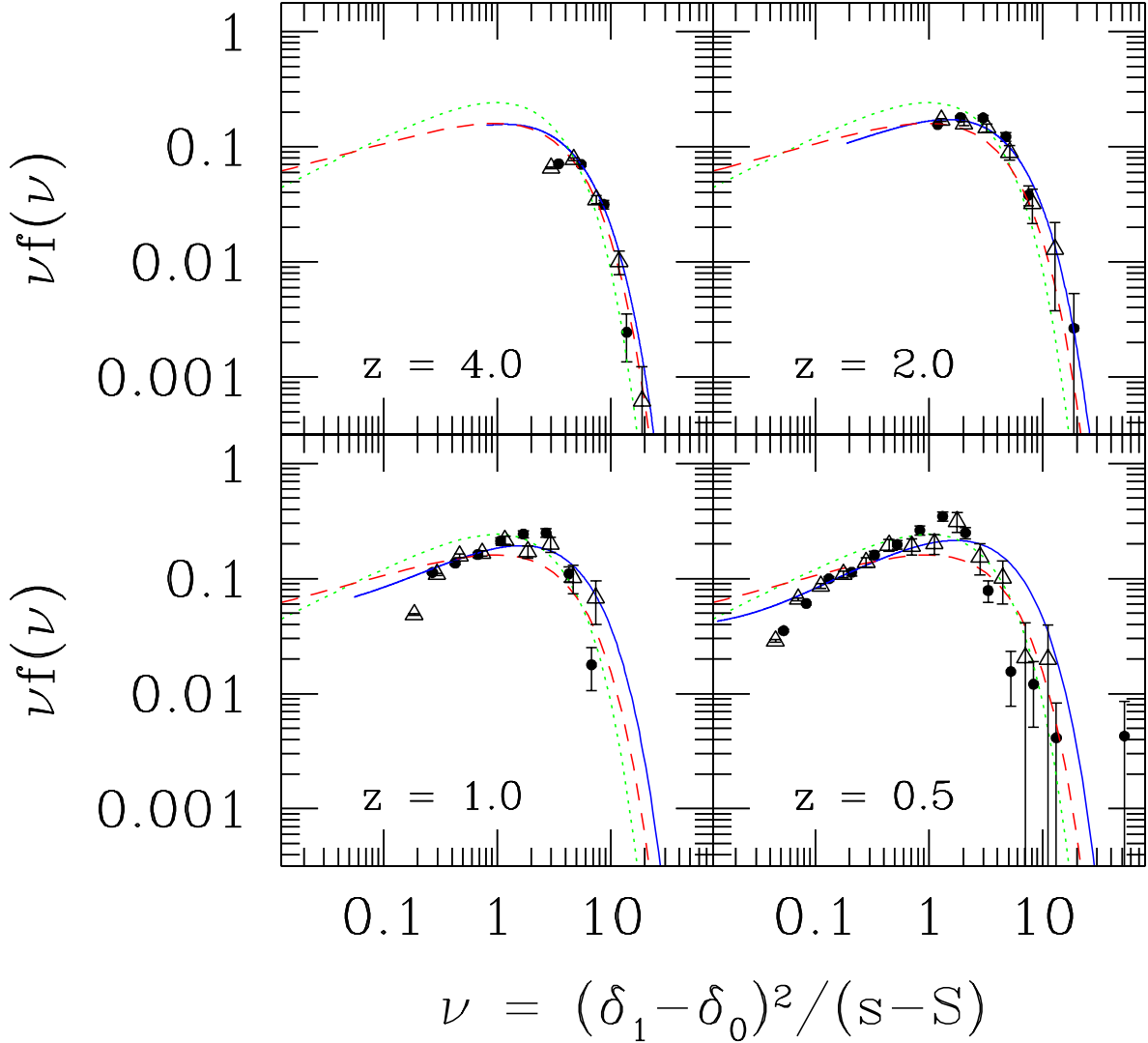}
\caption{Rescaled conditional mass functions in the SCDM model.  
Panels show the different redshift bins we studied earlier.  
Symbols in each panel show the different mass ranges we considered.  
Dotted line shows the constant barrier prediction (in these variables 
it is the same as the unconditional mass function), dashed line shows 
the result of rescaling the actual unconditional mass function, and 
solid curves show the result of rescaling the moving barrier predictions.}
\label{vfvscdm}
\end{figure*}

\begin{figure*}
\centering
\epsfxsize=\hsize\epsffile{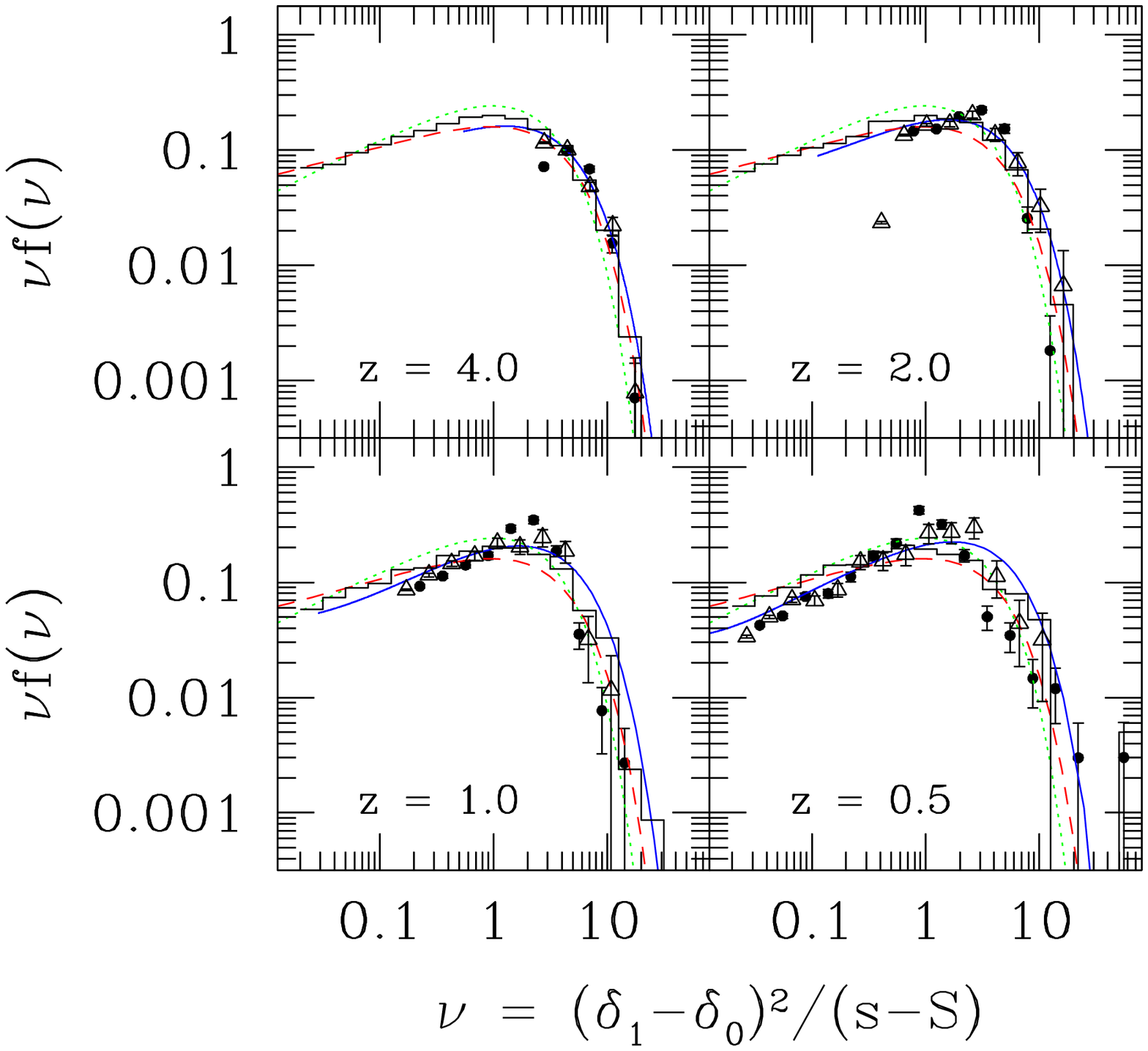}
\caption{Same as the previous figure, but for $\Lambda$CDM.}
\label{vfvlcdm}
\end{figure*}

The previous section we noted that, although the unconditional mass 
function is a universal function of $\nu$, this function is not the one 
predicted by the constant barrier model.  We argued that if we interpret 
the unconditional mass function as coming from a moving barrier, then we 
no longer expect the conditional mass function to be a universal function 
of $\nu$.  Figs.~\ref{vfvscdm} and~\ref{vfvlcdm} show this explicitly: 
they show the result of applying this rescaling to the conditional mass 
functions in the SCDM and the $\Lambda$CDM simulations we presented 
earlier.  

The four panels in each figure show the conditional mass function 
at each of the four redshifts we have been studying so far.  The symbols 
show the result of rescaling the conditional mass functions in the 
simulations for parent haloes with mass in the range 1-2 (filled circles) 
and 8-32 $M_*$ (open triangles) at $z=0$.  Notice that the symbols in 
each panel do not overlap exactly---at fixed $z$, the conditional 
mass functions for different parent haloes do not rescale exactly.  
In addition, the band traced out by the symbols at $z=4$ is quite 
different from the band traced out at $z=0.5$; the mass functions at 
different output times do not rescale either.  Both these findings 
illustrate our main point:  the conditional mass function is not a 
universal function of the scaling variable $\nu$.  

The dotted curves which are the same in all the panels show the 
predictions of the constant barrier model; they do not provide a 
good fit at any time for any mass range.  The dashed curves which 
are also the same in all the panels show the result of assuming that, 
upon rescaling, the conditional mass function will have the same 
shape as the unconditional mass function; although they provide a good 
fit at large $z$, they are increasingly in error at small $z$.  
This is true both for the SCDM and the $\Lambda$CDM models.  

The solid lines in the various panels show the predictions of our 
moving barrier model.  In this case, the predictions depend both on 
the parent mass range, and on the redshift at which the progenitors are 
identified:  we have chosen to show the predictions for the 1-2$M_*$ 
haloes only.  Whereas the model is in reasonable agreement with the 
simulations at large $z$, it has the wrong shape an small $z$.  This is 
not terribly surprising, because our formula for the first crossing 
distribution of the moving barrier model was only supposed to work at 
large $z$, but it is unfortunate that the disagreement at low $z$ is 
so bad!  One might have thought that the actual first crossing distribution 
may be in good agreement with the GIF simulations, and that it is only 
the analytic approximation which is in error at small $z$.  Unfortunately, 
this is not so.  The histograms in Fig.~\ref{vfvlcdm} show the result of 
simulating an ensemble of random walks to construct the conditional mass 
functions; although they are in slightly better agreement with the 
simulations, they are still quite different.  

The discrepancy between the simulations and our moving barrier model 
predictions are most pronounced when the subclump mass is $m/M \ge 1/2$.  
This suggests that our model is unable to describe the histories of 
clumps at small lookback times.  At small lookback times, one might 
worry that the spherical overdensity groupfinder we use to identify 
the subclumps in the simulations might find a different set of objects 
than a friends-of-friends algorithm.  Plots of the rescaled conditional 
mass function constructed using a friends-of-friends algorithm look 
very similar to the spherical overdensity results presented above---the 
discrepancy between model and simulations is independent of the choice 
of groupfinder.  

In addition, because the theory assumes that mass is conserved---all 
the mass of a subclump becomes part of the final halo---whereas this 
is not true in the simulations: some of the particles which make up 
the final object may have come from a subclump which merged along with 
most of its particles into a different object.
This means that there is some freedom associated with how we decide 
if a clump at an early time should be counted as a subclump of a halo 
at the final time.  We have tried two schemes for idenfifying subclumps:  
a progenitor is a clump which donates at least half its mass to the 
final object, or which donates at least one particle to the final object.  
Once we have made this decision, we must also decide what we wish to count 
as the mass of the parent object:  two natural choices are the mass at the 
final time (which, by definition is fixed for all earlier redshifts), or 
the mass which is got by summing up the masses of all the progenitor 
subclumps (which may depend on redshift).  The figures above are 
for the case in which a subclump is any clump which donates at least 
one particle to the final object, and the parent mass is defined as 
the mass at the final time (so it is independent of redshift).  
Although the actual conditional mass functions depend slightly on 
which combination of the above choices we make, the generic results 
shown above are independent of this choice.  

Before moving on, we think it worth noting that the discrepancies 
between the SCDM simulations and the dotted or dashed curves are 
qualitatively similar to the discrepancies in the $\Lambda$CDM case.  
This suggests that one should be able to find a model which can account 
for these discrepancies in a way which is independent of power-spectrum, 
redshift or cosmology.  Our moving barrier model is just not up to the 
task.  The next section studies why.  

\section{An extension}\label{sixd}
In the ellipsoidal collapse model envisaged by Bond \& Myers (1996) 
and implemented by Sheth, Mo \& Tormen (2001), the collapse of a patch 
is determined by the surrounding shear field.  
In a Gaussian random field, the field around a patch may differ 
from patch to patch.  Appendix~\ref{grfs} provides a simple prescription 
for choosing a set of patches which have the correct ensemble averaged 
properties---in essence, this requires studying the first crossing 
distribution of six-dimensional random walks.  

Because a six-dimensional walk is computationally expensive, 
rather than choosing the distribution of initial patches from this 
exact statistical distribution, Sheth, Mo \& Tormen suggested it should 
be a good approximation to use an appropriately chosen mean value, and 
neglect the scatter around this value.  This allowed them to reduce what 
is a six-dimensional random walk to a one-dimensional walk.  It is this 
one-dimensional walk which we have considered so far.  One might worry, 
however, that the discrepancy between model predictions and simulation 
results we found in the previous section may actually be due to our 
neglect of the scatter around the average value.  
The main purpose of this section is to study this possibility.  

\begin{figure}
\centering
\epsfxsize=\hsize\epsffile{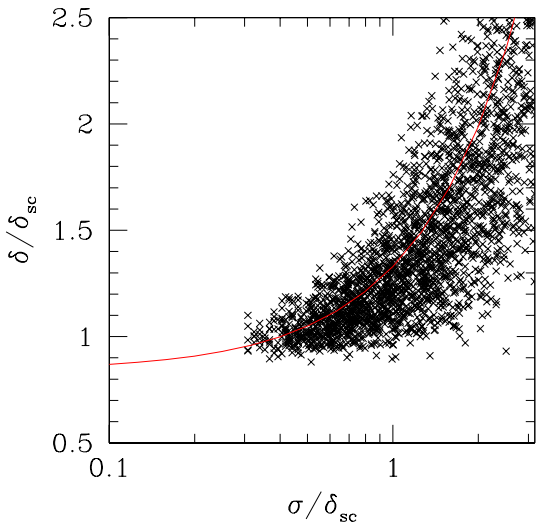}
\caption{Distribution of overdensities and scales at which the 
six-dimensional random walks crossed the ellipsoidal collapse 
barrier.  Solid curve shows the approximation used by 
Sheth, Mo \& Tormen (2001) in their, considerably simpler, 
one-dimensional random walks.  }
\label{scattersmt}
\end{figure}

\begin{figure}
\centering
\epsfxsize=\hsize\epsffile{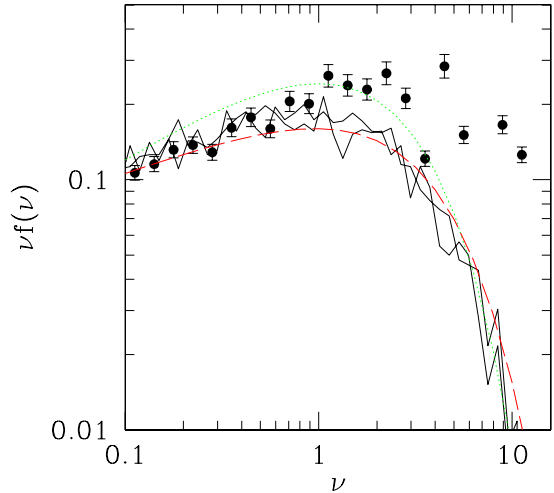}
\caption{Scaled unconditional (jagged curves) and conditional (filled 
circles) mass functions associated with six-dimensional random walks 
which cross the ellipsoidal collapse barrier.  Dashed line shows the 
unconditional mass function which fits the mass function of bound objects 
in simulations of clustering well.  As discussed in the previous section, 
the dashed curve is well described by the distribution of first crossings 
by one-dimensional random walks, of a barrier which is associated 
with ellipsoidal collapse.  Dotted line shows the corresponding 
spherical collapse prediction.  }
\label{walk6d}
\end{figure}

Fig.~\ref{scattersmt} shows the result of generating an ensemble of 
four thousand six-dimensional random walks associated with Gaussian 
random fields as described in the Appendix.  The walks are stopped 
when they cross the barrier associated with the ellipsoidal collapse 
model of Bond \& Myers (1996).  The crosses show the values of $\delta$ 
and $\sigma$ at which the six-dimensional walks crossed the ellipsoidal 
collapse barrier $\delta_{\rm crit}(e,p)$:  we actually used the simple 
fit, equation~3 in Sheth, Mo \& Tormen (2001), to the critical density 
required for collapse $\delta_{\rm crit}(e,p)$.  The solid curve 
shows the approximation used by Sheth, Mo \& Tormen; it provides 
a rather good description of how $\delta_{\rm crit}$ increases with 
increasing $\sigma$.  The Appendix describes the reason for this 
in more detail.  For now, we simply note that because the solid curve 
provides a reasonably good description of the crosses, the first 
crossing distributions of the six-dimensional walks considered here 
are unlikely to be very different from the first crossing distributions 
associated with the (considerably simpler) one-dimensional walks studied 
in the previous sections of this paper.

The two jagged curves in Fig.~\ref{walk6d} show this explicitly.  
They show the first crossing distributions associated with the 
six-dimensional walks which cross the $z=0$ and $z=0.5$ six-dimensional 
ellipsoidal collapse barriers.  They have been rescaled 
similarly to how the unconditional mass functions in simulations rescale:  
$\nu = (\delta_{\rm sc}/\sigma)^2$.  
After rescaling, the two curves appear similar, as they should; 
note that they are reasonably like the dashed curve, which shows 
equation~(\ref{giffit}), and they are rather different from the dotted 
curve which shows the spherical collapse prediction.  

Recall that the dashed curve is very similar to the mass function 
one gets by simulating one-dimensional random walks.  
Because the jagged curves are in quite good agreement with the 
dashed curve, the first crossing distributions associated 
with the six-dimensional walks are actually rather similar to those 
associated with the (considerably simpler) one-dimensional walks studied 
in the previous sections.  This agrees with what our conclusions from 
Fig.~\ref{scattersmt}.  The Appendix describes the reason for this in 
more detail.  

Having shown that the unconditional mass functions associated with 
the six-dimensional random walks are in reasonable with numerical 
simulations, we now turn to the conditional mass functions---the test 
which our one-dimensional random walks failed.  
The solid symbols with error bars show the conditional mass 
functions associated with the six-dimensional walks, expressed in the 
scaled units of the previous section:  
$\nu = 0.707\,\delta^2_{\rm sc}\,0.5^2/(s-S)$, where $s$ denotes the 
value of $\sigma^2$ at which the $z=0.5$ six-dimensional boundary was 
crossed, and $S$ is the scale on which the lower $z=0$ six-dimensional 
boundary was crossed.  
Note the excess of points at large values of $\nu$; these conditional 
mass functions are quite different from the conditional mass functions 
in simulations.  Indeed, the discrepancy between the excursion set 
predictions and the simulations of hierarchical clustering has got 
worse!  

There are at least two reasons why the excursion set approach with 
one-dimensional random walks may fail at small lookback times.  The 
first is the excursion set neglect of correlations between scales 
(Peacock \& Heavens 1990; Bond et al. 1991).  At large lookback times 
most subclumps are a small fraction of the mass of the parent halo, 
so the smoothing scale associated with the subclumps is sufficiently 
different from that of the parent that the neglect of correlations 
between the two scales is probably justified.  At smaller lookback 
times the parent and subclump scales are not so well separated, so the 
neglect of correlations is a more likely to be a bad approximation.  
This is one possible reason for the agreement at large lookback times 
despite the discrepancy at low redshift.  
The second possibility is that the one-dimensional parametrization of 
ellipsoidal collapse outlined by Sheth, Mo \& Tormen (2001) is too simple.  
The results of this section suggest that, in fact, it is the first 
possibility which is the cause of the discrepancy.  

\section{Discussion}\label{discuss}
Sheth, Mo \& Tormen (2001) argued that a simple modification to the 
original excursion set approach was enough to improve agreement 
between the predictions of the approach and numerical simulations.  
The modification they suggested was to the value of the linearly 
extrapolated critical overdensity $\delta_{\rm c}$ associated with 
the collapse of an object.  The spherical collapse model assumes that 
this value is independent of the mass of the collapsed object, whereas 
ellipsoidal collapse makes $\delta_{\rm c}$ depend on $m$.  In the 
context of the excursion set approach, this corresponds to studying 
the first crossing statistics of a set of moving, rather than constant 
barriers.  We also argued that a moving barrier also provides a simple 
way in which the excursion set appraoch can be extended to apply to 
models in which the initial dark matter distribution is not completely 
cold.  

Because moving barrier models are so useful, we provided analytic 
approximations for the required first crossing distributions 
(equations~\ref{taylors} and~\ref{fsS}), and showed that they were 
reasonably accurate.  
Although our formulae for the conditional mass functions (solid curves 
in Figs.~\ref{clcdm}--\ref{ndlcdm}) are slightly different from, and 
usually more accurate than, those one obtains by a simple rescaling of 
the unconditional mass function (dashed curves in the same figures), 
this simple rescaling of the unconditional mass function is still more 
accurate than what one gets if one rescales the constant barrier 
formulae (dotted curves).  
We showed that the predicted unconditional and conditional mass 
functions were in reasonably good agreement with results from 
numerical cosmological simulations (Figs.~\ref{clcdm}--\ref{ndlcdm}),
and we also provided a simple efficient algorithm which allows one 
to generating masses which have the correct universal mass function 
(Section~\ref{generate}).  

However, we showed that neither the constant nor the moving barrier 
models were able to describe the simulation results at small 
lookback times (Figs.~\ref{vfvscdm} and~\ref{vfvlcdm}).  This means 
that our results for the conditional mass functions cannot be used 
to generate realizations of the forest of merger history trees.  
We argued that this discrepancy was most likely due to the excursion 
set approach's neglect of correlations between scales 
(e.g. Peacock \& Heavens 1990; Bond et al. 1991; Monaco 1997b).  
While this neglect is a bad approximation at small lookback times, 
it is reasonably accurate at large lookback times.  This is why the 
excursion set approach is able to provide a reasonably good description 
of clustering at high redshift, even though it is inaccurate at small 
redshifts.  

\begin{figure}
\centering
\epsfxsize=\hsize\epsffile{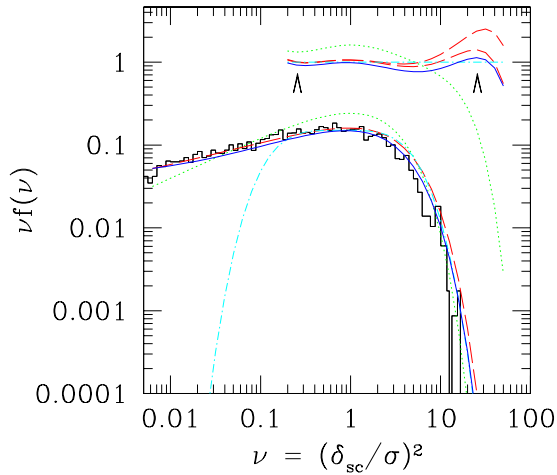}
\caption{Comparison of the various mass functions described in this 
paper, with the one which fits the Hubble volume simulations presented 
in Jenkins et al (2001).  Upper curves show the residuals between our 
analytic formulae and the Jenkins et al. fitting formula.  }
\label{hblvol}
\end{figure}

Before concluding, we will consider how some of our results are 
related to other work in the literature.  
Recently, Jenkins et al. (2001) showed that, although the mass 
functions in their simulations scaled in accordance with the excursion 
set prediction, our equation~(\ref{giffit}) slightly overestimated the 
unconditional mass functions in their simulations.  
We thought it would be interesting to show the various approximations
to the mass function which we presented in this paper on one plot.  
The dot-dashed curve in Fig.~\ref{hblvol}, with cutoffs at low and 
high masses shows the fitting function Jenkins et al. proposed, 
which fits their simulations well, the dashed curve shows the 
fitting formula of equation~(\ref{giffit}) with $a=0.707$ (following 
Sheth \& Tormen 1999), the histogram shows the distribution one gets by 
simulating random walks with the ellipsoidal collapse moving barrier 
(equation~\ref{bec}, following Sheth, Mo \& Tormen 2001), the solid 
curve shows our approximate formula for this first crossing distribution 
(equation~\ref{taylors}), and the dotted curve shows the spherical 
collapse, constant barrier prediction (Press \& Schechter 1974; 
Bond et al. 1991).  Simulations currently available do not probe the 
regime where $\nu \le 0.3$ or so (the Jeans mass is at about 
$\nu\approx 0.03$).  

\begin{figure}
\centering
\epsfxsize=\hsize\epsffile{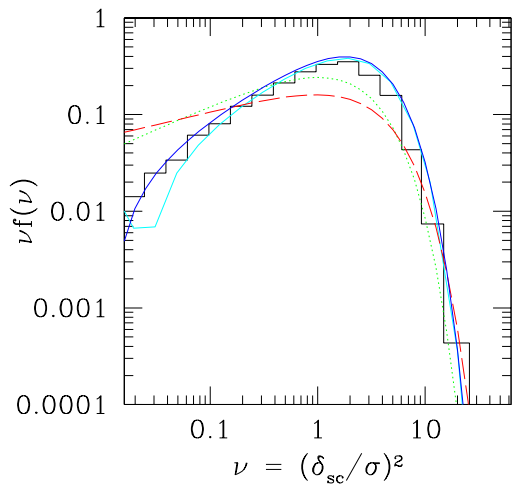}
\caption{The unconditional mass function associated with a model of 
collapse which was presented in Monaco (1997a,b).  Whereas the height of 
the barrier studied in our paper increases with distance, the one 
proposed by Monaco decreases with distance.  Despite the dramatic 
differences between the two barriers, inserting this shape into our 
equation~(\ref{taylors}) provides a good fit (solid curve) to the 
exact result (histogram).  }
\label{monaco}
\end{figure}

The upper set of curves show the residuals between our formulae 
and the one provided by Jenkins et al., in the regime to the right
of the low-mass cut-off (marked by an arrowhead).  In addition to 
the previously mentioned formulae, we have included a new dashed 
curve which shows the result of changing $a$ in equation~(\ref{giffit}) 
from 0.707 to 0.75.  This simple change appears to be all that is 
necessary to reduce the discrepancy between it and the simulations 
substantially.  Our formula differs dramatically from the one 
Jenkins et al. propose at small masses.  We hope that simulations 
in the near future will be able to address which low mass behaviour 
is correct.  

To illustrate that our formulae really do work for a large class of 
barrier shapes, Fig.~\ref{monaco} shows the first crossing 
distribution associated with the barrier discussed by Monaco (1997a,b):
\begin{displaymath}
B(S) = \delta_{\rm sc}\,
\Bigl[1.82/\delta_{\rm sc} - 0.69\,\sqrt{S/S_*}\Bigr],
\end{displaymath}  
where, as throughout this paper, $S_*\equiv\delta_{\rm sc}^2$.  
The height of this barrier decreases with $S$, so it really is quite 
different from ours (Sheth, Mo \& Tormen 2001 discuss the physical 
reason why).  (We chose not to present results for the barrier shape 
studied by Del Popolo \& Gambera (1998) because their shape is not 
so different from ours, whereas Monaco's really is quite different.  
The fact that this barrier decreases with $S$ means that all walks are 
guaranteed to cross it, and that there is no fragmentation associated 
with this barrier shape.  (This is in contrast to barriers whose height 
increases sufficiently strongly with $S$; for the Sheth, Mo \& Tormen 2001 
barrier shape studied in the main text, unbound mass and fragmentation 
are features which are formally possible but rare in practice.)  
The histogram shows the numerical Monte-Carlo first crossing distribution, 
and the two solid curves shows our analytic approximation, computed by 
inserting this barrier shape into our equation~(\ref{taylors}).  
The curve which provides a slightly worse fit to the histogram shows the 
result of using the first five terms in the series (as we did for the 
other figures in this paper); the other curve shows what happens if 
we use the first ten terms instead.  
Just for comparison, the dotted and dashed curves show the spherical 
collapse prediction, and the one which actually fits the cosmological 
simulations (equation~\ref{giffit}).  The figure shows that we are able 
to describe the first crossing distribution of this barrier shape well.  
This means that one could, in principle, use our formula, with Monaco's 
barrier, to study the conditional mass functions associated with his 
parametrization of nonlinear collapse---we have not pursued this 
further, although comparison of this first crossing distribution 
with the $z=0.5$ panels in Figs.~\ref{vfvscdm} and~\ref{vfvlcdm} suggest 
that this might be a useful exercise.  

In summary, we have provided a formula which describes the first 
crossing distribution of independent uncorrelated Brownian motion 
random walks, for a wide class of moving barriers.  This formula 
can be used to provide simple accurate predictions for a number of 
statistical quantities associated with the formation and clustering 
of dark matter haloes, all within the same formalism.

\bigskip

Many thanks to Lauro Moscardini for his realistic skepticism that 
this project would ever be completed!  We thank the TMR European 
Network ``The formation and evolution of galaxies'' under contract 
ERBFMRX-CT96-086 for financial support.  
RKS is supported by the DOE and NASA grant NAG 5-7092 at Fermilab.  
He thanks the Astronomy Department at the University of Padova for 
hospitality in May 2000.
The N-body simulations used in this paper are publically available at 
{\tt http://www.mpa-garching.mpg.de/NumCos}.  The simulations were 
carried out at the Computer Center of the Max-Planck Society in Garching 
and at the EPCC in Edinburgh, as part of the Virgo Consortium project, 
the members of which we would also like to thank.

\appendix

\section{The linear barrier}
Suppose that the barrier shape increases linearly with increasing 
variance $S=\sigma^2$:  
\begin{equation}
B(S,z) = \delta_{\rm c}(z)\,\Bigl(1 + S/S_*(z)\Bigr),
\label{linbar}
\end{equation}
where $\delta_{\rm c}(z)=\delta_{\rm c0}(1+z)$ and 
$S_*(z) = S_{*0}(1+z)^2$ if $\Omega=1$.  These scalings with $z$ 
are what is required by the self-similarity of Brownian motion.  
In the Bond et al. (1991) formulation of the constant 
barrier problem, it is customary to express the $S$ axis in the units 
it had at some fiducial time, say $z=0$, and to study the successive 
crossings of barriers having different values of $z$.  
In effect, the constant height Press--Schechter barrier has 
$S_{*0}=\infty$, so in that case only the scaling of the 
$y$-axis was apparent.  

This linear barrier shape is motivated by the observation that 
the GIF simulations have fewer low mass haloes relative to high 
mass ones, as compared to what is predicted by the constant barrier 
model.  This means that, at least for some range of $S$, the moving 
barrier must have a positive slope, since this would make it 
relatively easier to cross at small $S$ (large mass) than at large
$S$ (small mass), as compared to a barrier of constant height.  

An additional reason for considering this linear barrier is the 
following.  Recent work (Bode, Ostriker \& Turok 2001) suggests that 
the halo mass function in which the dark matter is warm initially 
has even fewer low mass halos than cold dark matter based 
ellipsoidal collapse models predict.  
In the context of the approach outlined by Sheth, Mo \& Tormen (2001), 
the physical reason for this is relatively simple:  low mass haloes 
do not form because they are hotter initially, so a larger overdensity 
is required to hold them together against the thermal pressure which 
prevents collapse, or against the stronger shearing from the velocity 
field.  This suggests that the critical density required for collapse 
by the present time, $\delta_{\rm ec}(m)$, should increase even more 
strongly with decreasing $m$ than it does when the dark matter is cold.  
The warm dark matter model is not particularly well motivated, 
and its free parameters have not yet been fixed, it seems premature to 
provide a detailed $\delta^{\rm WDM}_{\rm ec}(m)$ relation at the present 
time.  For this reason, the linear barrier considered in this appendix 
should be thought of as an example of what happes when the barrier height 
increases even more steeply with decreasing mass than it does in the 
Sheth, Mo \& Tormen cold dark matter models.  

\begin{figure*}
\centering
\mbox{\psfig{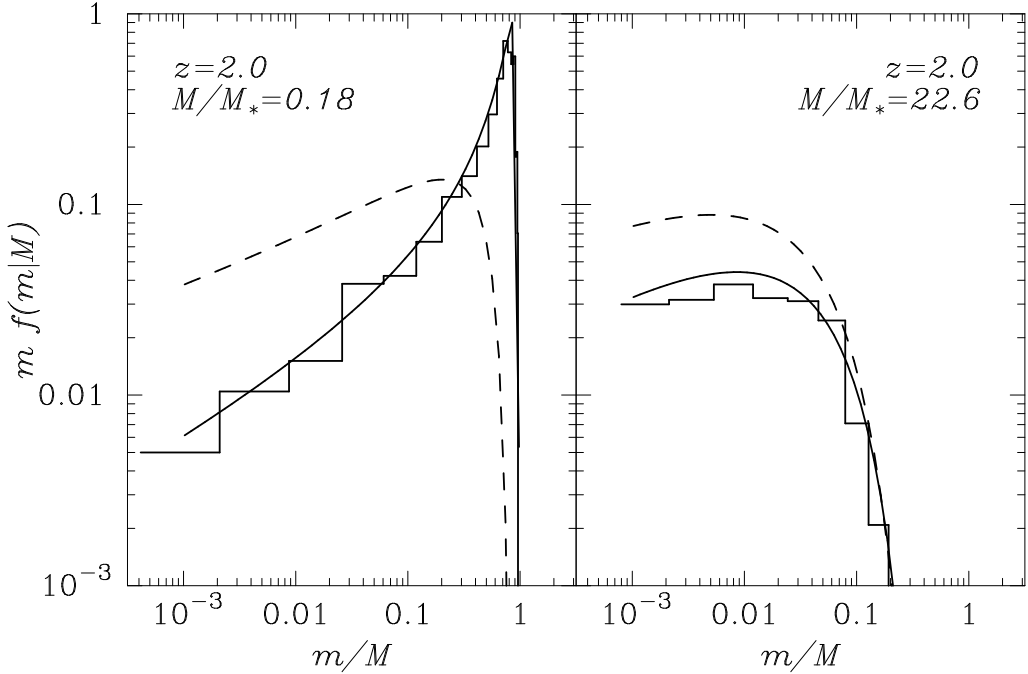}}
\caption{The conditional mass functions associated with the linear 
barrier.  Smooth solid curves show the analytic linear barrier 
prediction, dashed curves show the conditional constant barrier 
distribution for comparison, and histograms show the result of 
our numerical Monte--Carlo calculation.}
\label{lin2bar}
\end{figure*}

In the constant barrier model, all random walks were guaranteed to 
cross the barrier.  This is because the rms height of random walks 
at $S$ is proportional to $\sqrt{S}$, so at sufficiently large $S$, 
all walks will have crossed the constant barrier.  As a result, the 
associated first crossing distribution is normalized to unity:  since 
each random walk is associated with a volume element in the initial 
Lagrangian space, this is usually interpretted as meaning that, in 
the constant barrier model, all the mass in the universe is bound up 
in collapsed objects of some mass, however small.  In contrast, the 
linear barrier~(\ref{linbar}) increases to arbitrarily high 
values at high $S$.  Because the rms height of the random walk grows 
more slowly than the rate at which the barrier height increases, 
there is no guarantee that all random walk trajectories will intercept 
this barrier.  Indeed, only a fraction 
${\rm e}^{-2\delta^2_{\rm c}(z)/S_*(z)}$ of them will.  So, in the 
linear barrier model, not all initial volume elements are associated 
with bound haloes.  Since not all particles in numerical cosmological 
simulations are associated with bound haloes anyway (the fraction of 
unbound mass is typically on the order of $\sim$10\%, though how much 
of this is a consequence of limited resolution in the simulations is 
uncertain), this feature of the moving barrier model may or may not be 
a good thing.  In any case, this is one qualitative difference between 
the moving boundary model and a model with constant barrier height.  
(Readers who dislike this feature of the linear model are invited to 
patch a constant barrier of height $\delta_{\rm c0}$ at small $S$ to a 
linearly increasing one at intermediate $S$ to another constant one 
(but now at at a greater height, of course) at a large $S$ of their 
choice, and to compute the associated mass function!  This can be 
done relatively easily using the results presented below.)  

In addition to the question of normalization, barriers whose heights 
increase with $S$ more rapidly than $S^{0.5}$ will have mass functions 
in which the low mass end is depleted relative to the constant barrier 
case.  In the linear barrier model studied below, the mass function has 
an exponential cutoff at both low and high masses.  Because the barrier 
shape associated with ellipsoidal collapse grows only as $S^{0.6}$ 
both these effects are extremely weak.  Nevertheless, it is worth bearing 
these features in mind as bigger and better simulations become available.

The second important qualitative difference between a moving barrier 
model and the original constant barrier model is the following:  
whereas the $y$-intercept $B(0,z)$ increases as $z$ increases, 
the slope decreases as $(1+z)^{-1}$.  This means that it is possible 
for barriers to intersect at finite values of $S$.  
For example, two linear barriers $B(S,z_0)$ and $B(S,z_1)$, with 
$z_1>z_0$, will intersect at that critical value of $S$ at which 
$B(S,z_0)=B(S,z_1)$:  $S_{\rm 01}/S_{*0} = (1+z_0)(1+z_1)$.  
This means that all trajectories which first cross $B(z_0)$ 
at $S>S_{\rm 01}$ must have crossed $B(z_1)$ at a smaller 
value of $S$.  The logic of Lacey \& Cole (1993) then says 
that all  haloes at $z_0$ that are less massive than the associated 
critical mass $M_{01}$ were formed by fragmentation of a halo which, 
at $z_1>z_0$, was more massive.  

Again, this may or may not be a good thing.  Simulations show 
that $\sim$20\% of the total mass ever associated with progenitors 
of a halo does not find its way to the parent halo (Tormen 1998).  
Presumably this reflects the fact that while small haloes may 
fragment, the fragmented mass is not a very large fraction of the 
total mass.  Simulations also show that some subclumps pass through 
the virial radius of a given parent halo a number of times, 
each time depositing some fraction of their mass, before they 
finally become part of the parent halo.  It may be that a 
moving barrier model is able to incorporate and perhaps even 
quantify these effects.  

The first crossing distribution of trajectories that do cross the 
linear barrier is Inverse Gaussian:
\begin{equation}
f(S,z)\ {\rm d}S = 
{B(0,z)\over\sqrt{2\pi S}}\ 
\exp\left(-{B^2(S,z)\over 2S}\right)\,{{\rm d}S\over S} ,
\label{invg}
\end{equation}
(Whitmore 1978; also see Sheth 1998).  
The associated unconditional mass function is got by inserting this 
in equation~(\ref{fnunm}).  Whereas the mass function in the constant 
barrier case has an exponential cutoff at the high mass end, the mass 
function in this linear barrier model has exponential cutoffs at both 
low and high masses.  Recall this arises from the fact that barrier 
height increases so rapidly at large $S$.  Therefore, the mass function 
associated with this linear barrier has fewer small mass objects than 
the constant barrier predicts, in qualitative agreement with the GIF 
simulations.  (The agreement is certainly not quantitative, but the 
linear barrier is only used here for illustrative purposes.)

Since a linear barrier is linear whatever the origin of the coordinate 
system, the solution to the two barrier problem is also Inverse Gaussian.  
That is, given $z_1>z_0$ and given that the first crossing of $B(z_0)$ 
occured at $S_0\le S_{01}$, the probability that the first crossing 
of $B(z_1)$ occurs in the range ${\rm d}S_1$ about $S_1$ is 
given by equation~(\ref{invg}), with the substitutions 
$S\to (S_1-S_0)$, and $B\to B_{10}$, where 
\begin{displaymath}
B_{10}(S_1-S_0) = B(S_1,z_1) - B(S_0,z_0).  
\end{displaymath}
As was the case for the unconditional mass function, the height of 
this barrier diverges as $S_1-S_0\to\infty$, so not all trajectories 
intersect it.  Again, it seems reasonable to associate the fraction 
that do not with the fraction of the parent halo mass that is not 
associated with bound subclumps.  (Those readers who computed 
the mass functions associated with the patched
constant--linear--constant barriers may disregard the previous 
two sentences, provided they first compute the associated 
two-patched-barrier problem!)

The smooth solid curves in Fig.~\ref{lin2bar} show this 
conditional distribution for a few representative choices of the 
parent halo mass $M$.  The parents were assumed to have formed 
at $z_0=0$, and the progenitor distributions are shown at an 
earlier redshift $z_1$.  The underlying power spectrum was chosen 
to be the same as the GIF SCDM power spectrum.  It is also 
straightforward to solve this problem numerically:  
for the values of $M$ shown in the figures, the histograms show 
the distribution generated by simulating the crossings of the higher 
linear barrier by $10^3$ random Brownian motion trajectories that 
started at the initial positions $B(S,z=0)$, where $S(M)$ is given 
by the GIF power spectrum.  This figure has been included mainly 
to show that our Monte--Carlo code works, because we will use the 
code in the next section.  We have also verified that the numerical 
code gives the correct conditional and unconditional mass functions 
when the barrier heights are constant.  

For completeness, we also show the bias relations associated with 
this model.  
Following Mo \& White (1996) (also see Sheth \& Tormen 1999),
the mean Lagrangian bias between haloes and mass is 
\begin{equation}
\delta_{\rm h}^{\rm L}(1|0) = 
{f(S_1,z_1|S_0,z_0)\over f(S_1,z_1)} - 1.  
\label{biasl}
\end{equation}
The limit $M_0\gg M_1$ and $|\delta_0|\ll \delta_1$ 
is sometimes called the peak-background split.  
In this limit 
\begin{displaymath}
\delta_{\rm h}^{\rm L}(1|0)\to (\nu_1/\delta_1)\,\delta_0, \ \ \ 
{\rm where}\ \ \nu_1\equiv \delta^2_{\rm c}(z_1)/S_1
\end{displaymath}
for the linear barrier model.  
Massive haloes have small values of $S$, so they have large 
values of $\nu$.  Small haloes have $\nu\sim 0$.  
Thus, in this limit, less massive haloes are unbiased relative 
to the mass, whereas massive haloes are positively biased.  
For comparison, the corresponding limit for the constant barrier is
\begin{displaymath}
\delta_{\rm h}^{\rm L}(1|0)\to \delta_0\,(\nu_1 - 1)/\delta_1.  
\end{displaymath}
Thus, the predictions of the linear and constant barriers are 
similar for massive haloes, but they differ for less massive ones.  
In particular, less massive haloes in Lagrangian space are anti-biased
in the constant barrier model, whereas they are unbiased in the 
linear barrier model.  

The halo-to-mass bias in the evolved Eulerian space for the constant 
barrier can be computed by expanding 
\begin{equation}
\delta_{\rm h}^{\rm E}(1|0) = 
(1+\delta)\,\left[1 + \delta_{\rm h}^{\rm L}(1|0)\right] - 1
\label{biase}
\end{equation}
to lowest order in $\delta$ (Mo \& White 1996).  
In this limit, $\delta\approx\delta_0$, so 
$\delta_{\rm h}^{\rm E}(1|0) = (1 + [\nu_1 - 1]/\delta_1)\,\delta$ 
for the constant barrier.  The same logic gives 
$\delta_{\rm h}^{\rm E}(1|0) = (1 + \nu_1/\delta_1)\,\delta$ 
for the linear barrier.  Again, the predictions of the 
constant and linear barriers agree for massive haloes, but 
are different for less massive ones; the haloes in the linear 
barrier model are more positively biased.  This is encouraging, 
because, as mentioned in the introduction, this is in qualitative 
agreement with the trend seen in numerical simulations.  

The rate of increase of the ellipsoidal collapse barrier is shallower 
than for the linear barrier discussed here.  Since it is intermediate
between the linear barrier and the constant spherical collapse barrier, 
we might reasonably expect the large scale bias factor of less massive 
haloes in the ellipsoidal collapse model to be greater than that 
associated with the spherical collapse model.  Since the ellipsoidal 
collapse barrier rises less steeply than linear, the modified large 
scale bias will be somewhat less than the linear barrier prediction.  
Fig.~4 of Sheth, Mo \& Tormen (2001) shows that this is, indeed, the case.  

We argued at the start of this section that warm dark matter models 
can be parametrized by making the critical density for collapse 
depend more strongly on mass than when the dark matter is cold.  
In this respect, a comparison of the linear barrier formulae given 
here with the results for the barrier written down by 
Sheth, Mo \& Tormen (2001) shows why, generically, warm dark matter 
models are expected to have fewer low mass haloes and, consequently, 
different bias relations, particularly at the low mass end, than cold 
dark matter models.  This is in qualitative agreement with the 
numerical simulations of Bode, Ostriker \& Turok (2000).  

\section{Distribution of density and angular momentum of a patch in a Gaussian random field}\label{grfs}

Let $d_{ij}=\nabla_{ij}\phi$, where $\phi$ is the initial potential, 
denote the various components of the deformation tensor ${\cal D}$ 
(here $1\le i\le 3$ and similarly for $j$).  
Following, e.g. Bardeen et al. (1986), these components are 
\begin{eqnarray}
  d_{11} &=& (-y_1 - 3y_2/\sqrt{15} - y_3/\sqrt{5})/3 \nonumber\\
  d_{22} &=& (-y_1 + 2y_3/\sqrt{5})/3 \nonumber \\
  d_{33} &=& (-y_1 + 3y_2/\sqrt{15} - y_3/\sqrt{5})/3 \nonumber \\
  d_{12} &=& y_4/\sqrt{15} \nonumber \\
  d_{23} &=& y_5/\sqrt{15} \nonumber \\
  d_{13} &=& y_6/\sqrt{15}
\end{eqnarray}
where the $y_i$s are independent Gaussian variates with zero mean 
and variance $\sigma^2$.  Poisson's equation says that the trace of 
this matrix, Tr$({\cal D})$ equals the overdensity $\delta$.  Thus, 
\begin{equation}
\delta = d_{11} + d_{22} + d_{33} = -y_1;
\end{equation}
the final expression shows explicitly that $\delta$ is a Gaussian 
random variate.  

The eigenvalues of this matrix are the roots of the characteristic 
equation 
\begin{equation}
P(\lambda) = {\rm Det}[{\cal D} - \lambda {\cal I}] 
           = -a_0 - a_1 \lambda - a_2 \lambda^2 - \lambda^3,
\end{equation}
where ${\cal I}$ is the identity matrix, and the $a_k$s are various 
combinations of the $d_{ij}$s got by expanding the expression above 
and ordering by powers of $\lambda$.  
Since ${\cal D}$ is a $3\times 3$ real symmetric matrix, $P(\lambda)$ 
is a cubic with three real roots which satisfy 
\begin{eqnarray}
  \lambda_1 + \lambda_2 + \lambda_3  &=& - a_2 \nonumber \\
  \lambda_1 \lambda_2 + \lambda_2 \lambda_3 + \lambda_1 \lambda_3 &=& a_1 
   \nonumber \\
  \lambda_1 \lambda_2 \lambda_3 &=& -a_0.
\end{eqnarray}
Because rotations leave the trace unchanged, the overdensity $\delta$ 
is the sum of the three eigenvalues, so $-a_2 = \delta = -y_1$.  
In addition, the square of the angular momentum is proportional to 
\begin{eqnarray}
r^2 &=& {(\lambda_1-\lambda_2)^2\over 2} + {(\lambda_2-\lambda_3)^2\over 2} +
         {(\lambda_1-\lambda_3)^2\over 2} \nonumber \\
     &=& a_2^2 - 3 a_1
\end{eqnarray}
(e.g. Heavens \& Peacock 1988; Catelan \& Theuns 1998).  
This shows that to get $r^2$ we don't need to solve the cubic---we just 
need to read off the appropriate coefficients of the characteristic 
equation.  Thus, we find that 
\begin{eqnarray}
 r^2  &=& \delta^2 + 3 (d_{12}^2 + d_{13}^2 + d_{23}^2 
             - d_{11}d_{22} - d_{11}d_{33} - d_{22}d_{33}) \nonumber \\
      &=& (y_2^2 + y_3^2 + y_4^2 + y_5^2 + y_6^2)/5
\end{eqnarray}
Although the first line suggests that $r^2$ is coupled to $\delta$, 
the final expression shows that it is not.  In particular, the expressions 
above show that $\delta$ is distributed as a Gaussian, and $r^2$ is an 
independent variate drawn from a Chi-square distribution with 
five degrees of freedom, $\chi^2_5(\sigma)$.  The fact that the 
overdensity and the square of the angular momentum are independent 
does not seem to have been noticed before.  A $\chi^2_5(\sigma)$ 
distribution is rather similar in shape to a Lognormal, so this provides 
a simple way to see why spins of peaks in Gaussian random fields are 
also approximately lognormal (Heavens \& Peacock 1988).  

The results above can be used to generalize the excursion set algorithm 
studied in the main text.  Set $n=0$ and $y_i(n) =0$ for $1\le i\le 6$.  
Thereafter, at each step labeled by $n$, choose six, rather than 
one, independent Gaussian random variates $g_i$, each with variance $s$.  
For each $i=1,6$ set $y_i(n) = y_i(n-1) + g_i$, and use these to compute 
$\delta$ and $r^2$.  These can be used to give the values of the 
overdensity and the angular momentum for the scale on which the variance 
is $\sigma^2\propto ns$ (recall that $n$ is the number of steps taken).  
Now check to see if $\delta=-y_1(n)$ exceeds a critical value, say 
$\delta_{\rm crit}(\sigma,r^2)$.  
If it does, the six-dimensional walk stops at this scale.  
If not, the walk continues to smaller scales.  
Because each of the $g_i$s is chosen independently of the values of 
the $y_i$s or of $\sigma$, the walk takes independent steps in the 
six-dimensional space; it is in this sense that this algorithm 
generalizes the one-dimensional excursion set random walk studied 
in the main body of this paper.  

The analysis above shows clearly that the one-dimensional random walk 
approach of Sheth, Mo \& Tormen (2001) corresponds to the following 
approximation.  Replace the dependence of $\delta_{\rm crit}(\sigma,r^2)$ 
on the random variate $r^2$ by a dependence on its average value 
$\langle r^2\rangle \propto \sigma^2$.  This means that the critical 
density for collapse is a function of $\sigma$ alone, 
$\delta_{\rm crit}(\sigma)$.  As a result, the random walk in 
six-dimensions can be reduced to a walk in one-dimension only, 
thereby greatly reducing the complexity of the problem.  

Recently just such a six-dimensional random walk algorithm has been 
used by Chiueh \& Lee (2001), although they did not notice the 
considerable simplifications which follow from the algebra presented 
above.  They simulated an ensemble of six-dimensional random walks, and set 
the parameters of the barrier to be crossed by requiring that the 
resulting first crossing distribution give the unconditional mass 
function.  In particular, they showed that 
$\delta_{\rm crit} = 1.5 [1 + (2r^2/3)^2/0.15]^{0.15}$ provided 
a good fit to the required critical value of the overdensity.  

\begin{figure}
\epsfxsize=\hsize\epsffile{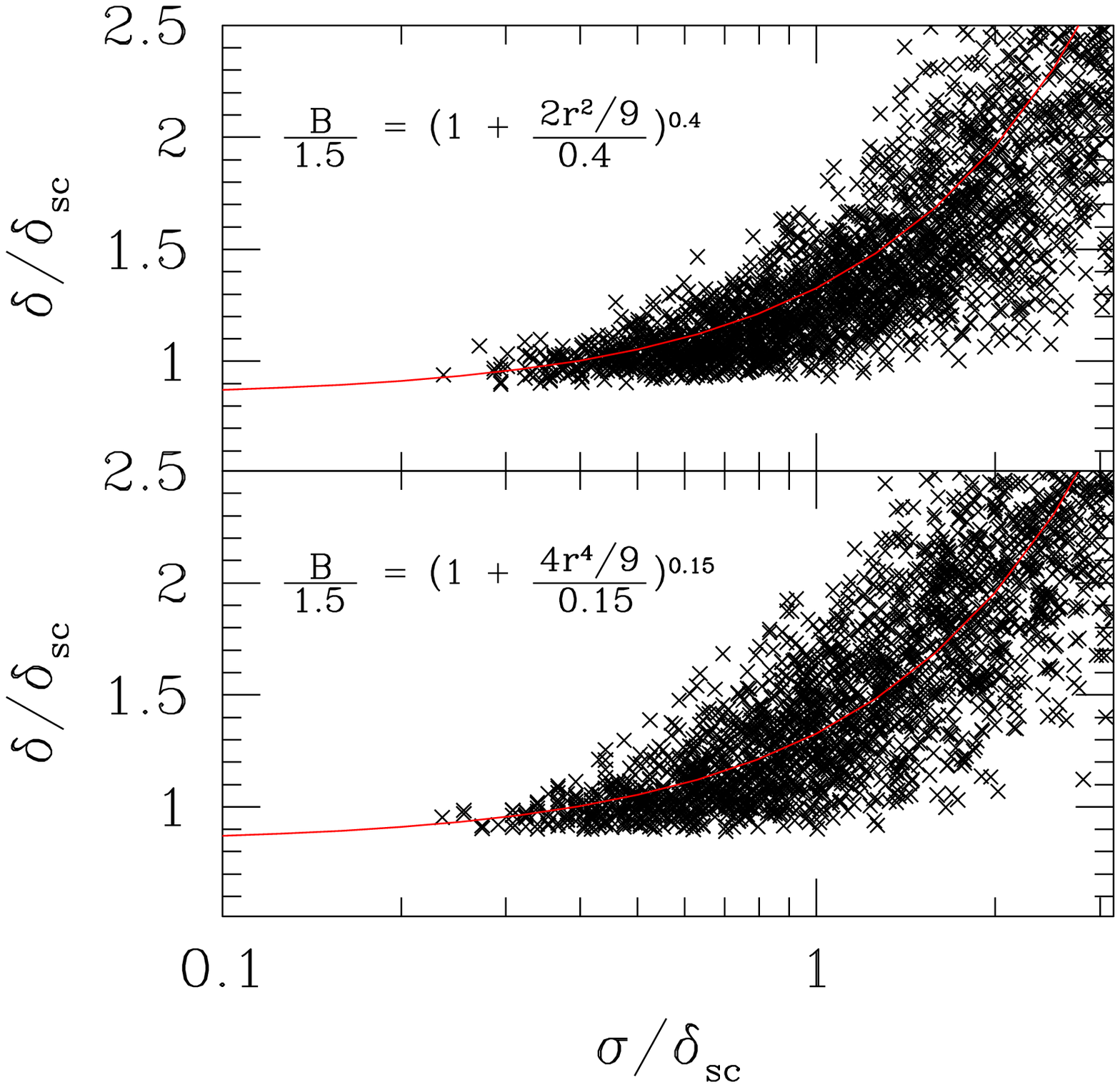}
\caption{Comparison of two prescriptions for collapse, both of which 
give good agreement with the unconditional mass function.  Crosses 
show the actual values of collapse densities and scales, and solid 
curves show the approximation assumed by Sheth, Mo \& Tormen (2001).}
\label{ecscatter}
\end{figure}

The algebra above allows one to see what such an approach implies.  
To do this, suppose the barrier is  $\delta_{\rm crit} = 1.686 (1 + r^2)$.  
The dependence on $r^2$ means that if the particle has walked 
to $\delta = 1.686$, it will still not have crossed, because $r^2$ 
is always certainly greater than zero.  So, to cross, the particle 
needs to have some $\delta > 1.686$.  How much greater?  
This depends on the typical value of $r^2$.  Because $r^2$ is drawn 
from a $\chi^2_5$ distribution, $\langle r^2\rangle \sim \sigma^2$.  
Now suppose that the $\chi^2_5$ distribution is very sharply peaked 
at its mean value (it quite well peaked, but taking the extreme 
case helps to see the argument).  This means that the barrier shape is 
something like   $\delta_{\rm crit} = 1.686 (1 + \sigma^2)$.  
The fact that a $\chi^2_5$ distribution is not very sharply peaked at 
its most probable value simply means that sometimes when 
$\delta = 1.686 (1 + \sigma^2)$ the particle will still be less than 
$\delta_{\rm crit}$, so the walk must go on.  
Of course, sometimes $r^2 < \sigma^2$, and in this case the walk will 
stop even if  $\delta < 1.686 (1 + \sigma^2)$.  
So, this means that we can think of the dependence on $r^2$ as making 
the critical value of the boundary height, when expressed as a 
function of $\sigma^2$ (the way Sheth, Mo \& Tormen 2001 did) a little 
fuzzy.  So, provided the $\chi^2_5$ distribution is not too broad, 
the considerably simpler one-dimensional random walk approach of 
Sheth, Mo \& Tormen (2001) should be a reasonable approximation.  

Fig.~\ref{ecscatter} shows the result of doing this for two choices of 
the barrier shape, both of which produce first crossing distributions 
which, when inserted into equation~(\ref{fnunm}), give mass functions 
of bound objects which are similar to the one in simulations of hierarchical 
clustering.  The upper panel was constructed using a barrier whose height 
increased linearly with $r^2$, and the lower panel shows results for the 
barrier shape used by Chiueh \& Lee (2001), which increases as $r^4$.  
The crosses show the values of $\delta$ and $\sigma$ at which each 
six-dimensional random walk crossed the barrier.  The solid curve shows 
the approximation used by Sheth, Mo \& Tormen (2001); it provides a 
reasonable description of the increase of $\delta_{\rm crit}$ with 
$\sigma$.  

Determining the barrier shape by requiring agreement with the 
clustering simulations is unsatisfying, especially in view of the 
fact that the two different boundaries given above provide equally 
adequate approximations to the mass function.  For this reason, 
the main text shows the result of combining the six-dimensional walk 
described above with the ellipsoidal collapse model of Bond \& Myers (1996).
This was relatively easy to do, because a simple fitting function for 
how the critical collapse boundary associated with this ellipsoidal 
collapse depends on the initial shear field has been given by 
Sheth, Mo \& Tormen (their equation~3 and their Fig.~1).  


\begin{thebibliography}{99}
\bibitem{wdm} Bode P., Ostriker J. P., Turok N., 2001, ApJ, to appear
\bibitem{bcek91} Bond J. R., Cole S., Efstathiou G., Kaiser N.,  1991, ApJ, 
379, 440
\bibitem{bm96} Bond J. R., Myers S., 1996, ApJS, 103, 1
\bibitem{cl01} Chiueh T., Lee J., 2001, ApJ, accepted
\bibitem{dpg98} Del Popolo A., Gambera M., 1998, AA, 337, 96
\bibitem{hp88} Heavens A. F., Peacock J. A., 1988, MNRAS, 232, 339
\bibitem{virgo} Jenkins A., Frenk C. S., White S. D. M., Colberg J. M., 
Cole S., Evrard A. E., Couchman H. M. P., Yoshida N., 2001, MNRAS, 321, 372
\bibitem{lc93} Lacey C., Cole S., 1993, MNRAS, 262, 627
\bibitem{lc94} Lacey C., Cole S., 1994, MNRAS, 271, 676
\bibitem{lk98}Lemson G., Kauffmann G., 1999, MNRAS, 302, 111
\bibitem{mw96} Mo H. J., White S. D. M., 1996, MNRAS, 282, 347
\bibitem{mjw97} Mo H. J., Jing Y., White S. D. M., 1997, MNRAS, 284,
189
\bibitem{plm97a} Monaco P., 1997a, MNRAS, 287, 753
\bibitem{plm97b} Monaco P., 1997b, MNRAS, 290, 439
\bibitem{ns99} Nusser A., Sheth R. K., 1999, MNRAS, 303, 685
\bibitem{ph90} Peacock J. A., Heavens A. F., 1990, MNRAS, 243, 133
\bibitem{ps74} Press W., Schechter P.,  1974, ApJ, 187, 425
\bibitem{rs96} Sheth R. K., 1996, MNRAS, 281, 1277
\bibitem{sp97} Sheth R. K., Pitman J., MNRAS, 289, 66
\bibitem{rs98} Sheth R. K., 1998, MNRAS, 300, 1057
\bibitem{sl98a} Sheth R. K., Lemson G., 1999a, MNRAS, 304, 767
\bibitem{sl98b} Sheth R. K., Lemson G., 1999b, MNRAS, 305, 946
\bibitem{st99} Sheth R. K., Tormen G., 1999, MNRAS, 308, 119
\bibitem{smt99} Sheth R. K., Mo H. J., Tormen G., 2001, MNRAS, 323, 1
\bibitem{bt97} Tormen G., 1997, MNRAS, 290, 411
\bibitem{bt98} Tormen G., 1998, MNRAS, 297, 648
\end{thebibliography}
\end{document}